# Kendall's Tau for Two-Sample Inference Problems

Yi-Cheng Tai[1], Weijing Wang[1], Martin T. Wells[2]
Correspondence: Weijing Wang, Institute of Statistics,
National Yang Ming Chiao Tung University, Hsin-Chu City, Taiwan, R.O.C.
Email: wjwang@stat.nctu.edu.tw

## Abstract

We consider a Kendall's tau measure between a binary group indicator and the continuous variable under investigation to develop a thorough two-sample comparison procedure. A zero association indicates that no one group is better than other, which is a broader condition than the equality of two survival (distribution) functions. In general, the Kendall's tau estimate provides an interpretable evidence measure of discrepancy between the two groups and hence serves as a useful alternative to the hazard ratio whose applicability depends on the proportional hazards assumption. For complete data, the Wilcoxon–Mann–Whitney statistic is an estimator of the Kendall's tau measure. For right censored data, we propose an inverse-probability-of-censoring-weighting estimator and show that it is a weighted log-rank statistic with weights that adapt to the censoring distributions. Theoretical properties of the derived estimators are investigated under the situations when the grouping is fixed in advance or is random. The large sample distribution theory results can be applied to test various hypotheses and construct confidence intervals based on Kendall's tau, which can quantify the effect size not provided by p values alone. An extensive simulation study examines the finite-sample properties of the proposed methodology and compares it to the log-rank and Gehan statistics. The proposed methodology is also applied to analyze several data examples.

---

[1] Institute of Statistics, National Yang Ming Chiao Tung University, Hsin-Chu City, Taiwan, R.O.C.
[2] Department of Statistics and Data Science, Cornell University, Ithaca, NY, United States.

## 1. INTRODUCTION

The two-sample comparison is a classical and important statistical problem. Most nonparametric approaches focus on hypothesis testing by specifying the null hypothesis as the equivalence of two distribution (survival or hazard) functions. Classical nonparametric test procedures are known to be sensitive to differences in the shapes of the distributions. For example, with complete data, the Wilcoxon–Mann–Whitney (WMW) statistic is one of such popular methods that is routinely used to compare two independent populations. However, several studies have shown that the WMW test to compare group means or medians is sensitive to deviations from the pure shift model (reviewed by Fagerland and Sandvik 2009); true significance levels sometimes below the nominal level, sometimes far above the nominal level. The pure shift model may not be very realistic for practical applications. In the right censored data setting, the log-rank (LR) test (Mantel, 1966) is most frequently applied and it is most powerful under the proportional hazard (PH) assumption. However, in real-world applications, it is common that survival curves cross, which in turn violates the PH assumption. In such cases, the power of the LR test may not be satisfactory. Weighted log-rank (WLR) tests (Harrington and Fleming, 1982) which incorporate weights to emphasize local differences have become alternative approaches in the case of non-proportional hazards. The Fleming and Harrington class of weights allows for tuning that depend on if one expects early, middle or late crossings, however, tuning those parameters is not straightforward since a misspecification may manifest in an even larger power loss than using the unweighted LR test (Jiménez et al. 2019). For example, the Gehan test (Gehan, 1965), which is an extended version of the WMW test for censored data, is a WLR test with the weight equal to the number at risk. The choice of weights for the WLR tests seems somewhat arbitrary and may lead to different, and possibly erroneous, conclusions.

It has become clear that statistical significance testing reported by p-values alone is not enough to make meaningful conclusions (Sterne, Cox and Smith, 2001). Commenting on this controversial topic, Cox emphasized that "statistical significance testing has a limited role, usually as a supplement to estimates". Hence, it is desirable to provide an interpretable measure which can assess treatment efficacy or quantify the discrepancy between two distributions. The hazard ratio is commonly chosen but it becomes less suitable when the PH assumption does not hold. For example, when a treatment has delayed effect, the PH assumption is no longer valid (Xu et al., 2017). The restricted mean survival time (RMST) is a popular alternative to hazard ratio (Royston and Parmar, 2013; Kim et al. 2017). Although the RMST is an interpretable measure, it is subject to the specification of a clinically meaningful time point by which the area under the survival curve is compared for the two groups.

In this article we study Kendall's tau estimators between a (fixed or random) binary group indicator and the continuous variable are compared. A zero association implies that no one group dominates the other, which includes the situations that the two distributions are identical or have crossing points but overall "balance out" the effect. The sign and magnitude of Kendall's tau yields an interpretable measure of the discrepancy between the two groups and hence can be used as an alternative to the hazard ratio. In randomized clinical trials, specifying a hypothesis usually involves the determination of a margin for equivalence or non-inferiority (Walker and Nowacki, 2011). The proposed Kendall's tau measure, which does not rely on the proportional hazard assumption, is an intuitive and reliable choice for setting the threshold value. For complete data fixed group assignment case, it is known that the WMW statistic is an estimator of Kendall's tau coefficient (Kendall, 1976). The case of random group indicators arises when it is not possible to assign treatments to fixed groups, perhaps for practical or ethical reasons. Often observational studies are collected from databases that contain outcomes with heterogeneous treatment exposures and causally-relevant confounders. We also propose an estimator of Kendall's tau suitable for censored data and its modified version if the tail information is missing due to censoring.

In this article, we develop theoretical properties of the Kendall's tau estimators under several sampling settings, which can be applied to perform hypothesis testing and construct confidence intervals. In Section 2, we derive large-sample properties of the WMW statistics based on complete data under the fixed and random grouping designs. In Section 3, we propose an inverse-probability-of-censoring-weighting (IPCW) estimator of the Kendall's tau statistic for right censored data and its modified version which imputes the order relationship in the tail area by parametric distributions. It is shown that the proposed estimator is a WLR statistic with weights that adapt to the censoring distributions. Given the counting process framework combined with the U-statistics structure, we derive large-sample properties of the proposed estimator including analytic variance formula under the two grouping settings. In Section 4, we present selected simulations studies which examine finite-sample performances of the proposed methods. In Section 5, we apply the proposed methodology to analyze two comprehensive datasets. Concluding remarks are given in Section 6. The proofs of the results, additional derivations and extensive simulation results, as well as two more data examples are all given in the Supporting Information.

## 2. PROPOSED METHODOLOGY FOR COMPLETE DATA

### 2.1 WMW statistics and Kendall's tau

Consider two populations indexed by 0 and 1. Let $T^{(\ell)}$ be a common continuous variable of interest for Group $\ell$ with the distribution and survival functions denoted as $F_\ell(t) = \Pr(T^{(\ell)} \leq t)$ and $S_\ell(t) = \Pr(T^{(\ell)} > t)$, respectively for $\ell = 0, 1$. Let $T_i^{(\ell)}$ ($i = 1, ..., N_\ell$) be a random sample of $T^{(\ell)}$ and assume that $T_i^{(0)}$ and $T_j^{(1)}$ are independent for all pairs

$(i, j)$. The following WMW statistic is commonly adopted for testing the hypothesis $H_0 : S_0 = S_1$:

$$U = \sum_{i=1}^{N_0} \sum_{j=1}^{N_1} \left\{ I(T_i^{(0)} < T_j^{(1)}) - I(T_i^{(0)} > T_j^{(1)}) \right\}. \tag{1}$$

We can pool the two populations and define $T$ and $X$ as the continuous variable and the group indicator taking values of 0 and 1, respectively. Therefore $F_\ell(t)$ and $S_\ell(t)$ can be written as $F_\ell(t) = \Pr(T \leq t \mid X = \ell) = 1 - S_\ell(t)$ ($\ell = 0,1$). Let $(T_i, X_i)$ and $(T_j, X_j)$ ($i \neq j$) be independent replications of $(T, X)$. Consider two versions of Kendall's tau (Kendall 1976) between $T$ and $X$ as follows:

$$\tau_a = E\left\{ sign(X_i - X_j)(T_i - T_j) \right\},$$

$$\tau_b = E\left\{ sign(X_i - X_j)(T_i - T_j) \mid X_i \neq X_j \right\},$$

where the function $sign(u)$ takes values of $1, 0$ and $-1$ if $u > 0$, $u = 0$ and $u < 0$, respectively. The second measure $\tau_b$, which makes adjustments for ties in the $X$'s to confine the comparison for pairs in different groups, that take values in $[-1,1]$. The value of $\tau_b$ reflects how much and in what direction the two groups differ from each other. For example, $\tau_b > 0$ implies that $T$ in Group 1 tends to be larger than $T$ in Group 0. Note that

$$\tau_b = E\left\{ S_1(T^{(0)}) - F_1(T^{(0)}) \right\} = \int_0^\infty \left\{ S_1(t) - F_1(t) \right\} dF_0(t).$$

When $S_0(t) = S_1(t)$ for all $t$, we get $\tau_b = 0$ but the converse may not be true. For example if $\Pr(T^{(1)} > T^{(0)}) = 0.5$, $\tau_b = 0$. In this article, we also construct confidence intervals for $\tau_b$ and testing $H_0^{tau} : \tau_b = \tau_0$, where $\tau_0$ is a hypothesized value. The special case $H_0^{tau} : \tau_b = 0$ is more general than $H_0: S_0 = S_1$ since the former allows for the crossing of two functions. In practical applications, it is of more interest to investigate the situation that $H_0$ and $H_0^{tau}$ are both false. In this case, a confidence interval of $\tau_b$ can measure how much and in what direction the two samples differ.

Let $(T_i, X_i)(i=1,...,n)$ denote a random sample of $(T, X)$. We can write $U$ in (1) as

$$U = \sum_{1 \leq i < j \leq n} \left[ sign\{(X_i - X_j)(T_i - T_j)\} \right], \qquad (2)$$

where $N_\ell = \sum_{i=1}^{n} I(X_i = \ell)$ $(\ell = 0,1)$ subject to $n = N_0 + N_1$. Hence $\tau_a$ and $\tau_b$ can be estimated by $\bar{\tau}_a = U / \binom{n}{2}$ and $\bar{\tau}_b = U / (N_0 N_1)$, respectively. The estimator $\bar{\tau}_b$ is unbiased but the derivations of its distributional properties depend on whether $N_\ell$ is fixed or random. When the group indicator $X$ is random, $N_\ell$ follows Binomial$(n, p_\ell)$, where $p_\ell = \Pr(X = \ell)$ $(\ell = 0,1)$. This occurs in observational data or some randomized clinical trials in which the treatment allocation follows a randomization scheme with a pre-specified value of $p_\ell$. However, many two-sample statistics implicitly assume that the group indicators $(X_1,...,X_n)$ and the group size $N_\ell$ are known (fixed) constants. In Sections 2.2 and 2.3, we derive asymptotic distributions of $\bar{\tau}_b$ under the two, fixed and random X, grouping designs and apply the results to construct confidence intervals of $\tau_b$ and test $H_0^{tau}: \tau_b = \tau_0$, where $\tau_0$ is a pre-specified value.

## 2.2 Asymptotic properties of $\bar{\tau}_b$ under the fixed grouping design

Theorem 1 states the large sample properties of $\bar{\tau}_b$ when the group indicators and the group sizes $N_\ell = n_\ell$ ($\ell = 0,1$) are fixed.

**Theorem 1**: *When the group indicators are fixed, the estimator $\bar{\tau}_b$ is a two-sample U-statistic for the parameter $\tau_b$ of degree (1,1). Assume $\frac{n_1}{n_0 + n_1} \to p_1$ as $n_0, n_1 \to \infty$, and $p_0 = 1 - p_1$. It follows that $\sqrt{n}(\bar{\tau}_b - \tau_b)$ converges to a mean-zero normal random variable with variance equal to*

$$n\sigma^2_{\bar{\tau}_{b,F}} = \frac{1}{p_0}\sigma^2_{0,1} + \frac{1}{p_1}\sigma^2_{1,0},$$

where $\sigma^2_{0,1} = E\left[\left\{S_1(T^{(0)}) - F_1(T^{(0)})\right\}^2\right] - \tau_b^2$ and $\sigma^2_{1,0} = E\left[\left\{F_0(T^{(1)}) - S_0(T^{(1)})\right\}^2\right] - \tau_b^2$.

The proof of Theorem 1, which utilizes standard techniques of two-sample U-statistics (Serfling 2009), is given in Web Appendix A1 of the Supporting Information.

The components of $\sigma^2_{\bar{\tau}_{b,F}}$ can be estimated separately. Specifically $\sigma^2_{0,1}$ and $\sigma^2_{1,0}$ can be estimated by the sample moments, denoted as $\hat{\sigma}^2_{0,1}$ and $\hat{\sigma}^2_{1,0}$, based on $\bar{F}_1(T_i^{(0)})$ ($i = 1,...,n_0$) and $\bar{S}_0(T_j^{(1)})$ ($j = 1,...,n_1$) respectively, with $S_\ell$ and $F_\ell$ estimated by

$$\bar{S}_\ell(t) = \sum_{i=1}^{n_\ell} I(T_i^{(\ell)} > t)/n_\ell \text{ and } \bar{F}_\ell = 1 - \bar{S}_\ell$$

for $\ell = 0,1$, respectively. Under the special case where $H_0: S_0(t) = S_1(t)$ ($\tau_b = 0$), it is easy to see that $\sigma^2_{1,0} = \sigma^2_{0,1} = \frac{1}{3}$. Hence $n\sigma^2_{\bar{\tau}_{b,F}}$ becomes $\frac{1}{3}\left(\frac{1}{p_0} + \frac{1}{p_1}\right) = \frac{1}{3p_0 p_1}$. This is a classical result for the WMW two-sample test (Hoeffding, 1948).

### 2.3 Asymptotic properties of $\bar{\tau}_b$ under the random grouping design

When $N_1 = n - N_0$ is random, we can write

$$\sqrt{n}(\bar{\tau}_b - \tau_b) = \sqrt{n}\left(\frac{U}{N_0 N_1} - \frac{U}{n^2 p_0 p_1}\right) + \sqrt{n}\left(\frac{U}{n^2 p_0 p_1} - \tau_b\right), \tag{3}$$

where $U$ is expressed in terms of equation (2). The randomness of $N_1$ affects the distribution of the first component in (3) and its covariance with the second component. Properties of the second component in (3) can be directly obtained from $\sqrt{n}(\bar{\tau}_a - \tau_a)$ since

$$\sqrt{n}\left(\frac{U}{n^2 p_0 p_1} - \tau_b\right) = \frac{1}{2 p_0 p_1}\sqrt{n}(\bar{\tau}_a - \tau_a) + o_P(1).$$

The asymptotic properties of $\sqrt{n}(\bar{\tau}_a - \tau_a)$ are given in Web Appendix A2 of the Supporting Information which can be proved by directly applying asymptotic theory for U-statistics. Theorem 2 states the asymptotic properties of $\sqrt{n}(\bar{\tau}_b - \tau_b)$ under the random grouping design. The involved derivations are summarized in Web Appendix A3 of the Supporting Information.

***Theorem 2***: *When the group indicators are random and assume that $0 < p_1 = \Pr(X = 1) < 1$ and $p_0 = 1 - p_1$, $\sqrt{n}(\bar{\tau}_b - \tau_b)$ converges to a mean-zero normal random variable with variance equal to*

$$n\sigma^2_{\bar{\tau}_{b,R}} = \frac{\sigma_1^2}{(p_0 p_1)^2} - \frac{(p_1 - p_0)^2}{p_0 p_1}\tau_b^2,$$

where $V_0 = I(X = 0)\{S_1(T) - F_1(T)\}$, $V_1 = I(X = 1)\{F_0(T) - S_0(T)\}$ and

$$\sigma_1^2 = E\left\{(p_0 V_1 + p_1 V_0)^2\right\} - \tau_a^2.$$

The components of $\sigma^2_{\bar{\tau}_{b,R}}$ can be estimated separately as follows. Specifically, $p_\ell$ can be estimated by its empirical counterpart $\hat{p}_\ell = N_\ell / n$ ($\ell = 0, 1$) and $\sigma_1^2$ can be estimated by

$$\hat{\sigma}_1^2 = \sum_{i=1}^{n}\left(\hat{p}_0 V_{1,i} + \hat{p}_1 \hat{V}_{0,i}\right)^2 / n - \bar{\tau}_a^2,$$

where $\hat{V}_{0,i} = I(X_i = 0)\{\bar{S}_1(T_i) - \bar{F}_1(T_i)\}$, $\hat{V}_{1,i} = I(X_i = 1)\{\bar{F}_0(T_i) - \bar{S}_0(T_i)\}$ and

$$\bar{F}_\ell(t) = 1 - \bar{S}_\ell(t) = \sum_{i=1}^{n} I(T_i \leq t, X_i = \ell) / \sum_{i=1}^{n} I(X_i = \ell).$$

In the special case that $X$ and $T$ are independent, it follows that $\tau_b = 0, \sigma_1^2 = (p_0 p_1)/3$. Hence $n\sigma_{\bar{\tau}_{b,R}}^2 = 1/(3p_0 p_1)$ which is the same as $n\sigma_{\bar{\tau}_{b,F}}^2$ under the special case $H_0: S_0 = S_1$.

## 3. PROPOSED METHODOLOGY FOR RIGHT CENSORED DATA

### 3.1 Proposed estimator of $\tau_b$ and WLR statistics

Let $C^{(\ell)}$ be the censoring variable for Group $\ell$ with the survival function $G_\ell(t) = \Pr(C^{(\ell)} > t)$ ($\ell = 0,1$). Two-sample right censored data can be denoted as $\left(Y_i^{(\ell)}, \delta_i^{(\ell)}\right)$ ($i = 1,...,N_\ell$), where $Y_i^{(\ell)} = T_i^{(\ell)} \wedge C_i^{(\ell)}$ and $\delta_i^{(\ell)} = I\left(T_i^{(\ell)} \leq C_i^{(\ell)}\right)$ for $\ell = 0,1$. Let $\left(C_1^{(\ell)},...,C_{N_\ell}^{(\ell)}\right)$ be a random sample of $C^{(\ell)}$ and assume that $C_i^{(\ell)}$ is dependent of $T_i^{(\ell)}$ for all $i = 1,...,N_\ell$. To modify $U$ in (1) in the presence of right censoring, the value of $I(T_i^{(0)} < T_j^{(1)})$ is known only if $\delta_i^{(0)} = 1$ and the value of $I(T_i^{(0)} > T_j^{(1)})$ is known only if $\delta_j^{(1)} = 1$. Applying the "inverse-probability-censoring-weighting" (IPCW) technique, define

$$\hat{U} = \sum_{i=1}^{N_0} \sum_{j=1}^{N_1} \frac{\tilde{O}_{ij} \left\{I\left(Y_i^{(0)} < Y_j^{(1)}\right) - I\left(Y_i^{(0)} > Y_j^{(1)}\right)\right\}}{\hat{G}_0\left(Y_i^{(0)} \wedge Y_j^{(1)}\right) \hat{G}_1\left(Y_i^{(0)} \wedge Y_j^{(1)}\right)}, \quad (4)$$

where

$$\tilde{O}_{ij} = I\left(T_i^{(0)} \wedge T_j^{(1)} \leq C_i^{(0)} \wedge C_j^{(1)}\right) = I(Y_i^{(0)} < Y_j^{(1)}, \delta_i^{(0)} = 1) + I(Y_i^{(0)} > Y_j^{(1)}, \delta_j^{(1)} = 1)$$

and $\hat{G}_\ell(t)$ is the Kaplan-Meier estimator of $G_\ell(t)$ using $\left(Y_i^{(\ell)}, 1 - \delta_i^{(\ell)}\right)$ ($i = 1,...,N_\ell; \ell = 0,1$).

We can also develop the proposed statistic based on the pooled sample. Under the combined population, consider $(T, C, X)$ such that $T$ is subject to censoring by $C$ with $G_\ell(t) = \Pr(C > t \mid X = \ell)$ ($\ell = 0,1$). Observed variables are denoted as $(Y, \delta, X)$, where $Y = T \wedge C$, $\delta = I(T \leq C)$ and $X$ is a binary group indicator taking values of 0 and 1. The observed data can therefore be written as the triple $(Y_i, \delta_i, X_i)$ $(i = 1, ..., n)$. The proposed statistic $\hat{U}$ in (4) can be written as

$$\hat{U} = \sum_{1 \leq i < j \leq n} \frac{O_{ij} \, sign\{(X_i - X_j)(Y_i - Y_j)\}}{\hat{G}_0(\tilde{Y}_{ij}) \hat{G}_1(\tilde{Y}_{ij})}, \qquad (5)$$

where $\tilde{T}_{ij} = T_i \wedge T_j$, $\tilde{C}_{ij} = C_i \wedge C_j$, $\tilde{Y}_{ij} = Y_i \wedge Y_j$ and

$$O_{ij} = I(\tilde{T}_{ij} < \tilde{C}_{ij}) = I(Y_i < Y_j, \delta_i = 1) + I(Y_j < Y_i, \delta_j = 1).$$

We propose to estimate $\tau_b$ by $\hat{\tau}_b = \hat{U} / (N_0 N_1)$, where $\hat{U}$ is defined in (4) or (5). When the two groups have the same censoring distributions with $G_0(t) = G_1(t) = G(t)$, we set $\hat{G}_0(t) = \hat{G}_1(t) = \hat{G}(t)$ which is the Kaplan-Meier estimator of $G(t)$ based on the pooled sample $(Y_i, 1 - \delta_i)$ $(i = 1, ..., n)$. Note that the IPCW technique has been adopted by Lakhal et al. (2009) to estimate Kendall's tau under bivariate censoring.

In Web Appendices B1-B3 of the Supporting Information, we express $\hat{\tau}_b$ together with the LR and Gehan statistics under the same structures since the three statistics are weighted versions of each other. We show that $\hat{\tau}_b$ is a WLR statistic expressed in (S2) of the Supporting Information such that

$$\hat{\tau}_b = \frac{1}{N_0 N_1} \int_0^\infty \frac{R_0(u) R_1(u)}{\hat{G}_0(u) \hat{G}_1(u)} \{d\hat{\Lambda}_0(u) - d\hat{\Lambda}_1(u)\},$$

where $R_\ell(t) = \sum_{i=1}^n I(Y_i \geq t, X_i = \ell)$ and $\hat{\Lambda}_\ell$ is the estimate of the cumulative hazard function of $T \mid X = \ell$ $(\ell = 0, 1)$. Note that the weight function of $\hat{\tau}_b$ is free of the censoring distributions in the limit. We notice that $\hat{\tau}_b$ coincides with the statistic $-\int_0^\infty \hat{S}_1(u-)d\hat{S}_0(u)$ proposed by Efron (1967) for testing $H_0: S_0(t) = S_1(t)$, where $\hat{S}_\ell$ is the Kaplan-Meier estimator of $S_\ell$ ($\ell = 0,1$). It is worthy to mention that here we emphasize the role of $\hat{\tau}_b$ as an estimator of $\tau_b$ so that we derive the distributional properties of $\hat{\tau}_b$ under general situations including $\tau_b \neq 0$ which can be applied to a much broader class of inference problems.

**3.2 Asymptotic properties of $\hat{\tau}_b$ under fixed and random grouping designs**

Define the support points of $T$, $C \mid X = \ell$ and $Y$ as $\xi_T = \sup\{t : \Pr(T \geq t) > 0\}$, $\xi_{C_\ell} = \sup\{t : \Pr(C \geq t \mid X = \ell) > 0\}$ for $\ell = 0,1$ and $\xi_Y = \sup\{t : \Pr(Y \geq t) > 0\}$, respectively. Here we assume that $\xi_T \leq \xi_{C_0} \wedge \xi_{C_1}$ under which $\tau_b$ is identifiable based on censored data. The asymptotic properties of $\hat{\tau}_b = \hat{U}/(n_0 n_1)$, where $\hat{U}$ is defined in (4) and $N_\ell = n_\ell$ ($\ell = 0,1$) under the fixed grouping design, are stated in Theorem 3.

*Theorem 3: Assume that the group indicators are fixed and $\dfrac{n_1}{n_0 + n_1} \to p_1$ as $n_0, n_1 \to \infty$, $0 < p_1 < 1$, $\xi_T \leq \xi_{C_0} \wedge \xi_{C_1}$. If $E\left[\{\psi(\mathbf{Y}_j^{(1)}; \mathbf{Y}_i^{(0)})\}^2\right] < \infty$, where $\mathbf{Y}_i^{(\ell)} = (Y_i^{(\ell)}, \delta_i^{(\ell)})$ ($\ell = 0,1$), and*

$$\psi(\mathbf{Y}_j^{(1)}; \mathbf{Y}_i^{(0)}) = \frac{\tilde{O}_{ij}\{I(Y_i^{(0)} < Y_j^{(1)}) - I(Y_i^{(0)} < Y_j^{(1)})\}}{G_0(Y_i^{(0)} \wedge Y_j^{(1)})G_1(Y_i^{(0)} \wedge Y_j^{(1)})}$$

*is a two-sample kernel of order $(1,1)$. Then $\sqrt{n}(\hat{\tau}_b - \tau_b)$ converges to a mean-zero normal random variable with variance $n\sigma^2_{\hat{\tau}_{b,F}} < \infty$, where the variance formula is given in Web Appendix C1 of the Supporting Information.*

Note that $\xi_T \leq \xi_{C_0} \wedge \xi_{C_1}$ implies that $G_\ell(T_i^{(0)} \wedge T_j^{(1)}) > 0$ where $(T_i^{(0)}, T_j^{(1)})$ are a pair randomly chosen from $(T^{(0)}, T^{(1)})$. Hence $\psi$ is a bounded function which guarantees $E\left[\{\psi(\mathbf{Y}_j^{(1)}; \mathbf{Y}_i^{(0)})\}^2\right] < \infty$. Theorem 4 states the results for the random design and the derivations are given in Web Appendix C2 of the Supporting Information.

**Theorem 4:** *Assume* $0 < p_1 < 1$, $\xi_Y = \xi_T \leq \xi_{C_0} \wedge \xi_{C_1}$ *and* $E\{\zeta^2(\mathbf{Y}_i, \mathbf{Y}_j)\} < \infty$ $(i \neq j)$, *where* $\mathbf{Y}_i = (X_i, Y_i, \delta_i)$ *and*

$$\zeta(\mathbf{Y}_i, \mathbf{Y}_j) = \frac{O_{ij} \operatorname{sign}\{(X_i - X_j)(Y_i - Y_j)\}}{G_0(Y_i \wedge Y_j) G_1(Y_i \wedge Y_j)} \text{ is a symmetric kernel of order 2.}$$

*Then* $\sqrt{n}(\hat{\tau}_b - \tau_b)$ *converges to a mean-zero normal random variable with variance* $n\sigma^2_{\hat{\tau}_{b,R}} < \infty$, *where the variance formula and its estimator are given in Web Appendix C2 of the Supporting Information.*

In Web Appendix C3 of the Supporting Information, we show that under $H_0^{tau}: \tau_b = 0$ and $H_0: S_0 = S_1$, $(\hat{\tau}_b - \tau_b)/\sigma_{\hat{\tau}_{b,F}}$ and $(\hat{\tau}_b - \tau_b)/\sigma_{\hat{\tau}_{b,R}}$ are asymptotically equivalent. This means that the grouping mechanism has no effect from the perspective of hypothesis testing.

## 3.3 Modifications for Insufficient Follow-Up

In practical applications for the right censored data methods, it may happen that $\xi_T > \xi_{C_0} \wedge \xi_{C_1} = \xi_Y$ due to insufficient follow-up. To handle this issue under bivariate censoring, Fan et al. (2000) and Lakhal et al. (2009) suggest to use restricted versions of the original estimators. Applying their ideas to our setting, we can instead estimate $\tau_b^R$ defined as

$$\Pr\left((i,j) \text{ concordant} \mid X_i \neq X_j, \tilde{T}_{ij} \leq \xi_Y\right) - \Pr\left((i,j) \text{ discordant} \mid X_i \neq X_j, \tilde{T}_{ij} \leq \xi_Y\right),$$

which can be estimated by

$$\hat{\tau}_b^R = \frac{\hat{\tau}_b}{1 - \hat{S}_0(Y_{(n)})\hat{S}_1(Y_{(n)})}. \tag{6}$$

We now propose an alternative approach by interpolation such that the missing concordant information is imputed under selected parametric models. Let $t^*$ be a pre-specified constant satisfying $\Pr(Y > t^*) > 0$. Define

$$\hat{\tau}_{b,1}^* = \frac{1}{N_0 N_1} \sum_{i<j} \frac{O_{ij} I(\tilde{Y}_{ij} \leq t^*) sign(X_i - X_j)(Y_i - Y_j)}{\hat{G}_0(\tilde{Y}_{ij})\hat{G}_1(\tilde{Y}_{ij})}, \tag{7}$$

which is an estimator of

$$\tau_{b,1}^* = \Pr\big((i,j) \text{ concordant}, \tilde{T}_{ij} \leq t^* \mid X_i \neq X_j\big) - \Pr\big((i,j) \text{ discordant}, \tilde{T}_{ij} \leq t^* \mid X_i \neq X_j\big).$$

Furthermore, define the population value of the remaining component:

$$\tau_{b,2}^* = \Pr\big((i,j) \text{ concordant}, \tilde{T}_{ij} > t^* \mid X_i \neq X_j\big) - \Pr\big((i,j) \text{ discordant}, \tilde{T}_{ij} > t^* \mid X_i \neq X_j\big)$$
$$= \int_{t^*}^{\infty} S_1(t) dF_0(t) - \int_{t^*}^{\infty} S_0(t) dF_1(t).$$

The proposed modified estimator of $\tau_b$ can be written as $\hat{\tau}_b^* = \hat{\tau}_{b,1}^* + \hat{\tau}_{b,2}^*$, where

$$\hat{\tau}_{b,2}^* = \int_{t^*}^{\infty} \hat{S}_1^*(t) d\hat{F}_0^*(t) - \int_{t^*}^{\infty} \hat{S}_0^*(t) d\hat{F}_1^*(t), \tag{8}$$

$\hat{S}_\ell^*$ and $\hat{F}_\ell^* = 1 - \hat{S}_\ell^*$ ($\ell = 0, 1$) are parametric estimators of the marginal survival functions obtained from maximum likelihood estimation. Goodness-of-fit tests can be performed to select a best fitted parametric model. Sensitivity analysis can also be conducted by comparing $\hat{\tau}_b^*$ under different model choices.

Due to the complexity involved in analytic derivations, we will adopt the bootstrap approach for variance estimation. Under the random grouping design, a bootstrap sample is drawn from the combined sample; while under the fixed grouping design, separate bootstrap samples are drawn from the sub-datasets with $X = 1$ and $X = 0$ independently. By repeating the bootstrap resampling procedure many times, the approximate distribution of $\hat{\tau}_b^*$ can be obtained and used to estimate $\sigma_{\hat{\tau}_b^*}^2$. If the resulting intervals do not lie in $[-1,1]$, we can conduct further normalization to fix this problem. Specifically, we can replace $\hat{\tau}_{b,1}^*$ by

$$\tilde{\tau}_{b,1}^* = \left\{ \sum_{i<j} \frac{O_{ij} I(\tilde{Y}_{ij} \leq t^*)}{\hat{G}_0(\tilde{Y}_{ij}) \hat{G}_1(\tilde{Y}_{ij})} \right\}^{-1} \left\{ \sum_{i<j} \frac{O_{ij} I(\tilde{Y}_{ij} \leq t^*) sign(X_i - X_j)(Y_i - Y_j)}{\hat{G}_0(\tilde{Y}_{ij}) \hat{G}_1(\tilde{Y}_{ij})} \right\} \times Pr(\tilde{T}_{ij} \leq t^* \mid X_i \neq X_j),$$

where $Pr(\tilde{T}_{ij} \leq t^* \mid X_i \neq X_j) = 1 - \hat{S}_0(t^*)\hat{S}_1(t^*)$ and use $\tilde{\tau}_b^* = \tilde{\tau}_{b,1}^* + \hat{\tau}_{b,2}^*$ to estimate $\tau_b$.

## 4. SIMULATION STUDIES

Here we present selected simulation results based on the random grouping design. Additional simulation results are provided in Web Appendix D of the Supporting Information. We generate $(X_i, T_i)$ by first generating $X_i \sim Ber(p_1)$; and then $T_i \mid X_i = \ell$ and $C_i \mid X_i = \ell$ for $i = 1,...,n$. The Weibull distribution is adopted to model $T \mid X = \ell$ with the density function:

$$f_\ell(t) = \frac{\kappa_\ell}{\lambda_\ell} \left( \frac{t}{\lambda_\ell} \right)^{\kappa_\ell - 1} \exp\left\{ -\left( \frac{t}{\lambda_\ell} \right)^{\kappa_\ell} \right\} \quad (\kappa_\ell, \lambda_\ell > 0; \ell = 0,1).$$

Under the situation that $\xi_T \leq \xi_{C_0} \wedge \xi_{C_1}$, $C \mid X = \ell$ follows an exponential distribution with the hazard rate $\lambda_\ell^C$ for $\ell = 0,1$. If $\xi_T > \xi_{C_0} \wedge \xi_{C_1}$, the censoring variables are generated from the uniform distributions.

First, we examine the performances of $\bar{\tau}_b$, $\hat{\tau}_b$ and the corresponding confidence intervals based on the normality results of $(\bar{\tau}_b - \tau_b)/\hat{\sigma}_{\bar{\tau}_{b,R}}$ and $(\hat{\tau}_b - \tau_b)/\hat{\sigma}_{\hat{\tau}_{b,R}}$ stated in Theorems 2 and 4. In the computation of $\hat{\tau}_b$, we set $\lambda_0^C = \lambda_1^C = 1$ but estimate $G_0$ and $G_1$ separately. Table 1 summarizes the results based on $n = 400$. Similar analysis based on different sample sizes under both grouping designs are given in Web Tables 1-5 of the Supporting Information. Web Table 6 provides the results under unequal censoring with $\lambda_0^C = 1$ and $\lambda_1^C = 0.5$. From Table 1, we see that the coverage probabilities are close to the nominal level. This implies that the proposed variance formulae are correct and the normality approximation works well with $n = 400$ even under moderate or heavy censoring. Notice that the lengths of confidence intervals become wider when the censoring rates increase. When $\tau_b = 0$, the average lengths of the confidence intervals tend to be widest under the same level of $p_1$ in the three settings.

Then we evaluate the powers of the LR, Gehan and proposed $\hat{\tau}_b$ tests under the proportional hazard assumption such that the failure time in Group $\ell$ follows an exponential distribution with the hazard rate $\lambda_\ell$ for $\ell = 0,1$. Fixing $\lambda_0 = 1$, the value of $\lambda_1$ varies which also affects the value of $\tau_b$. We present the powers defined as the probabilities of rejecting $H_0: S_0 = S_1$ under different levels of $\tau_b$ and three censoring settings: $(\lambda_0^C, \lambda_1^C) = (1,1)$, $(0.2, 0.2)$ and $(0.05, 0.05)$ indexed by (a), (b) and (c) respectively. In Figure 1(a), in which the group-specific censoring proportions are highest, the power curve of the proposed $\hat{\tau}_b$ test almost overlaps with that of the LR test which is the optimal test under the PH assumption.

When the censoring rates decrease as shown in cases (b) and (c) of Figure 1, the power curve of the proposed test gets closer to that of the Gehan test since the two tests reduce to the WMW test in absence of censoring. The detailed information used to produce Figure 1 is summarized in Web Table 7 of the Supporting Information.

We further compare the three test statistics when the proportional hazard assumption does not hold. The two failure times follow the Weibull distributions with $(\kappa_0, \lambda_0) = (2, 1.2)$ and $(\kappa_1, \lambda_1) = (0.5, 2)$. Web Figure 1 of the Supporting Information shows the survival, density and hazard functions of the two distributions. Note that the hazard functions of the two groups behave in opposite directions. Since the two survival functions cross at the median points, this case corresponds to $\tau_b = 0$ but $S_0 \neq S_1$. In Figure 2, we plot the histograms of standardized values of the LR, Gehan and proposed $\hat{\tau}_b$ statistics based on 2000 runs. The results are compared under three censoring settings, where $\Pr(\delta = 0 | X = \ell)$ ($\ell = 0, 1$) are (a): 0.604, 0.560; (b): 0.490, 0.506; and (c): 0.390, 0.453. We see that the censoring distributions greatly influence the LR test since the peak of the histogram varies from negative to positive values. Under Case (a) where the censoring is heavier, the sign of the LR statistic indicates that the hazard of Group 0 is smaller but gives the opposite conclusion in Case (c) where the censoring is lighter. The Gehan test which assigns larger weights to early time points gives the consistent conclusion that Group 0 has a smaller hazard rate but it does not detect later difference of the two hazards. In contrast, the proposed $\hat{\tau}_b$ statistic is quite robust with respect to the censoring distributions. Summary information producing the histograms in Figure 2 is given in Web Table 8 of the Supporting Information.

Finally, we examine the performances of the proposed modified estimator when $\xi_T > \xi_{C_0} \wedge \xi_{C_1}$. We evaluate $\hat{\tau}_b^R$ and several versions of $\hat{\tau}_b^*$ under selected parametric models. The sample size is $n = 200$ and the number of replications is 2000. We generate $X \sim Ber(p_1 = 0.5)$ and $T \mid X = \ell$ which follows $Weibull(\kappa_\ell, \lambda_\ell)$ with $(\kappa_0, \lambda_0) = (0.5, 2)$ and $(\kappa_1, \lambda_1) = (2, 1.2)$ so that $\tau_b \approx 0$. The censoring variable $C \mid X = \ell$ follows $Uniform(0,1)$ for both $\ell = 0, 1$ so that the censoring rates are $\Pr(\delta = 0 \mid X = 0) = 0.6313$, $\Pr(\delta = 0 \mid X = 1) = 0.8099$ and $\Pr(\delta = 0) = 0.7210$. From Table 2, the performances of $\hat{\tau}_b^R$ and $\hat{\tau}_b^*$ interpolated under exponential models become worse since they don't recognize that the concordance patterns are different in early and late stages. The estimated values of $\hat{\tau}_b^*$ under the true Weibull model performs well as expected. However under a mis-specified log-normal model, the bias of $\hat{\tau}_b^*$ is also small. Although the information beyond $\xi_Y$ is not identifiable, the proposed extrapolation approach can be used as a tool for sensitivity analysis.

## 5. DATA ANALYSIS EXAMPLES

In this section, we apply the proposed methodology to analyze complete and censored data. Additional data analysis is given in Web Appendix E of the Supporting Information.

### 5.1 Comparison of Soil Water Contents in Two Fields

The soil water contents (% water by volume) collected from two experimental fields growing bell peppers are under comparison (Gumpertz et al., 1997). Field 1 (Group 1) contains 72 observations with the average water content around 11.42%; Field 2 (Group 0) contains 80 observations with the average water content around 10.65%. In Figure 3, the two survival curves have two crossing points. Given that $\bar{\tau}_b = 0.19$, the soil water content in Field 1 tends to be a little larger than that in Field 2. We reject $H_0^{tau}: \tau_b = 0$ if $\left|(\bar{\tau}_b - 0)/\hat{\sigma}_{\bar{\tau}_b}\right| > 1.96$, where $\hat{\sigma}_{\bar{\tau}_b}^2$ is a suitable estimate of $Var(\bar{\tau}_b)$. The suggested choice of $\hat{\sigma}_{\bar{\tau}_b}^2$ is $\hat{\sigma}_{\bar{\tau}_b}^2 = \hat{\sigma}_{\bar{\tau}_{b,F}}^2 = 0.00972$ (p-value = 0.0539) under the fixed grouping design; and $\hat{\sigma}_{\bar{\tau}_b}^2 = \hat{\sigma}_{\bar{\tau}_{b,F}}^2 = 0.00972$ (p-value = 0.0539) under the random grouping design. If for testing $H_0: S_0 = S_1$, we use $\hat{\sigma}_{\bar{\tau}_b}^2 = \dfrac{n}{3n_0 n_1} = 0.0088$ (p-value = 0.0427). The results based on $\bar{\tau}_b$ yield the conclusions of rejecting $H_0: S_0 = S_1$ and but not rejecting $H_0^{tau}: \tau_b = 0$. However, the p-value of the LR test is 0.2 which leads to not rejecting $H_0: S_0 = S_1$. The choice of variance formula depends on the how the data are collected. Nevertheless, the results of using different formula are very close although the p-values are around 0.05 leading to different conclusions in this example.

**5.2 Infection Time Comparison for Patients Receiving Two Catheter Placements**

From a dataset obtained from Klein and Moeschberger (2003) a study of 119 patients who received kidney dialysis were classified into two groups according to the way of placing the catheter. Group 0 contained 76 patients who utilized a percutaneous placement of their catheter; Group 1 contains 43 patients who utilized a surgical placement of their catheter. The variable under comparison is the time to infection. The overall censoring rate is 78% and the censoring rates in Groups 0 and 1 are 86% and 65%, respectively. The upper plot in Figure 4 indicate that $\hat{S}_0(t)$ and $\hat{S}_1(t)$ have an early crossing point around $t = 10$ (months) and both do not decrease to zero. From the lower plot of Figure 4, we see that $\hat{G}_0(t) < \hat{G}_1(t)$, a situation of unequal censorship.

The original estimate of $\tau_b$ is $\hat{\tau}_b = -0.44$ but the assumption $\xi_T \leq \xi_{C_0} \wedge \xi_{C_1}$ is obviously violated. To handle the missing concordant information in the tail region, we compute $\hat{\tau}_b^R$ in (6) and four versions of $\hat{\tau}_b^*$ interpolated by the tails of different parametric models beyond $t^* = y_{(n)} = 28.5$. The bootstrap percentile method based on the fixed grouping design is adopted to construct confidence intervals. The estimates of $\hat{\tau}_b^*$ and the corresponding 95% confidence intervals (in parenthesis) based on the Exponential, Weibull, log-normal and logistic models are $-0.508$ $(-0.779, -0.127)$, $-0.619$ $(-0.902, -0.224)$, $-0.650$ $(-0.920, -0.272)$ and $-0.484$ $(-0.754, -0.106)$, respectively. All the results show that patients in Group 0 using percutaneous placements tend to take longer time to develop infection than those in Group 1 using surgical placements. Notice that zero is not included in all the intervals. The proposed test based on $\hat{\tau}_b^* / \hat{\sigma}_{\hat{\tau}_b^*}$ shows strong evidence to reject $H_0^{\text{tau}} : \tau_b = 0$. However, the p-values of the LR and Gehan tests for testing $H_0 : S_0 = S_1$, are 0.1 and 0.42, respectively which means that the two existing tests fail to detect the difference between the two groups. We explain the differences of the three tests using the U-statistics expressions in Web Appendix B3 of the Support Information. From $\hat{S}_0(t)$ and $\hat{S}_1(t)$ in the upper plot of Figure 4, we see that most orderable pairs with large values of $\tilde{Y}_{ij}$ are associated with $sign(X_i - X_j)(Y_i - Y_j) = -1$. From the weight functions $\gamma_{ij}^L$, $\gamma_{ij}^G$ and $\gamma_{ij}^P$ in (S3) of the Supporting Information, Gehan statistic assigns equal weights to all pairs and the LR statistic assigns higher weight for larger $\tilde{Y}_{ij}$ but not as much as the proposed test which gives $\hat{\tau}_b = -0.44$. Recall that the proposed test is least vulnerable to the effect of censoring.

## 6. CONCLUDING REMARKS

For two-sample comparison, there are multiple reasons to support using estimates of $\tau_b$ as a preferred alternative inferential statistical procedure to the currently applied tests. First, the estimates of $\tau_b$ sign and magnitude provide direct and interpretable information about the magnitude of how much and in what direction the two groups differ. Next, significant tests based on the interpretable measure $\tau_b$ are more general and practical than simply testing the equality of the two survival functions especially in the setting when the assumption of proportional hazards is violated. We established the asymptotic distributions of $\bar{\tau}_b$ and $\hat{\tau}_b$, including analytic variance formulae, under both the fixed and random grouping designs based on complete or censored data. In particular, we showed that the proposed statistic $\hat{\tau}_b$, the LR and Gehan statistics are weighted versions of each other. The simulation analysis shows that the proposed $\hat{\tau}_b$ test is robust to the censoring distributions and the proportional hazard assumption. It has satisfactory power close to the LR test under the PH assumption when the censoring rate is high and reduces to the WMW test when there is no censoring.

In some medical applications, there is time lag before the treatment becomes effective. Statistical tests which ignore delayed treatment effect may lead to power loss. Weighted LR statistics with the weight appropriately chosen can increase the power, specification of the lag model is required to derive a suitable weight. For example, Xu et al. (2017) consider the standard delayed treatment effect scenario such that $\lambda_0(t) = \lambda_1(t)$ for $t \leq t_0$ and $\lambda_0(t) = \rho \lambda_1(t)$ for $t > t_0$, where $t_0$ denotes the hazard ratio changing point which is pre-specified. In Web Appendix D2, we modify our method to adjust for the problem of delayed treatment effect and perform simulations to examine the effect on the power. Compared with the modified WLR test proposed by Xu et. al (2017), our approach does not require the PH assumption after $t_0$. Since the proposed test statistic is also a WLR test, it is straightforward to extend it to the case of K-sample comparison. The theoretical structures can be further utilized to extend the analysis to more complicated settings.

## ACKNOWLEDGEMENT

Wang's research was supported by Ministry of Science and Technology, Taiwan grant 106-2118-M-009-002-MY3 and 109-2118-M-009-001-MY2. Wells' research was partially supported by National Institutes of Health awards and R01GM135926 1P01-AI159402.

The implementation code for statistical procedures developed in this article and some of the examples are available online. The *R* source code, the example data sets in Section 5 and some supplementary information can be downloaded at the GitHub site:

https://github.com/s07308/Two-Sample-Inference-with-Tau.

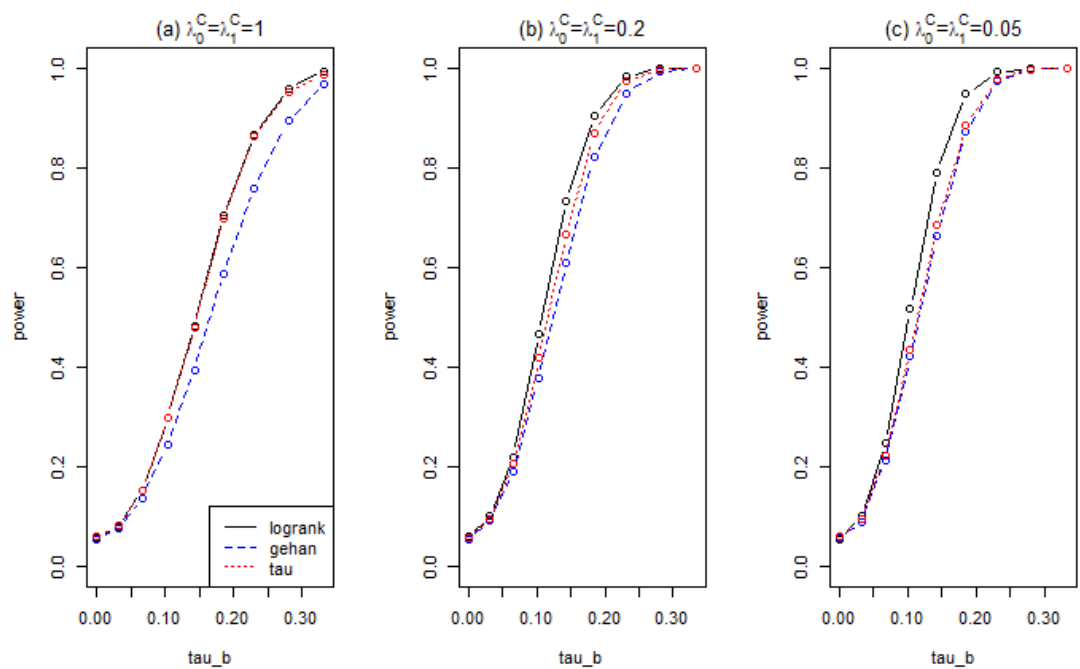

Figure 1: Power functions for the LR (Solid line), Gehan (dashed line) and proposed $\hat{\tau}_b$ (dotted line) tests for testing $H_0 : S_0 = S_1$ under the PH assumption and three configurations of the censoring distributions.

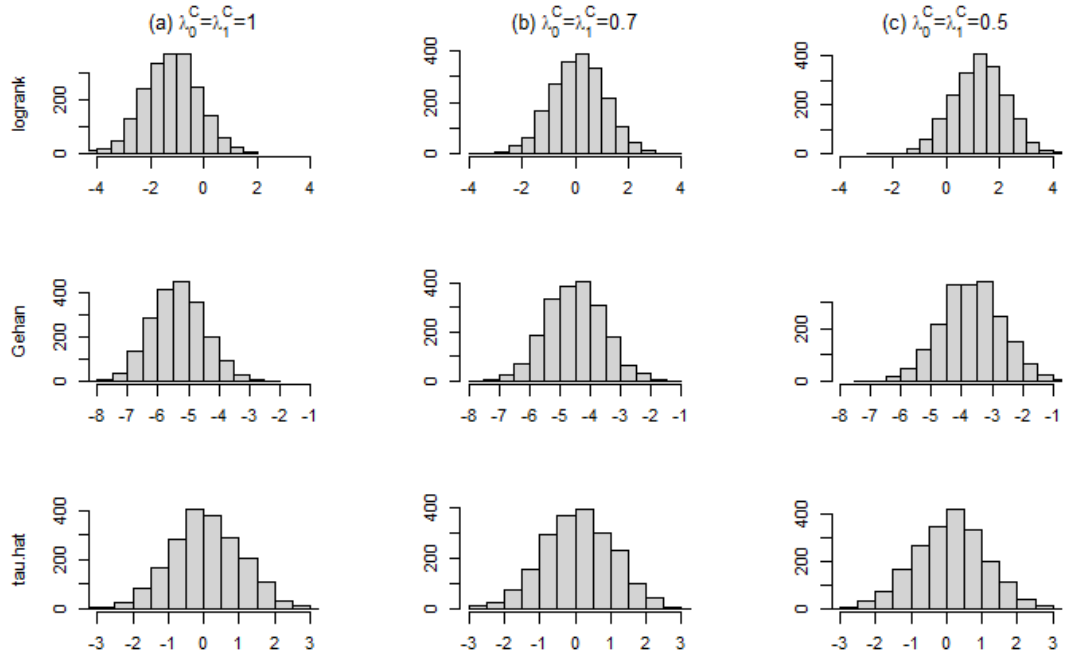

Figure 2: Histograms of the standardized LR, Gehan and proposed $\hat{\tau}_b$ statistics when the two Weibull survival functions have a crossing point with $S_0 \neq S_1$ and $\tau_b = 0$.

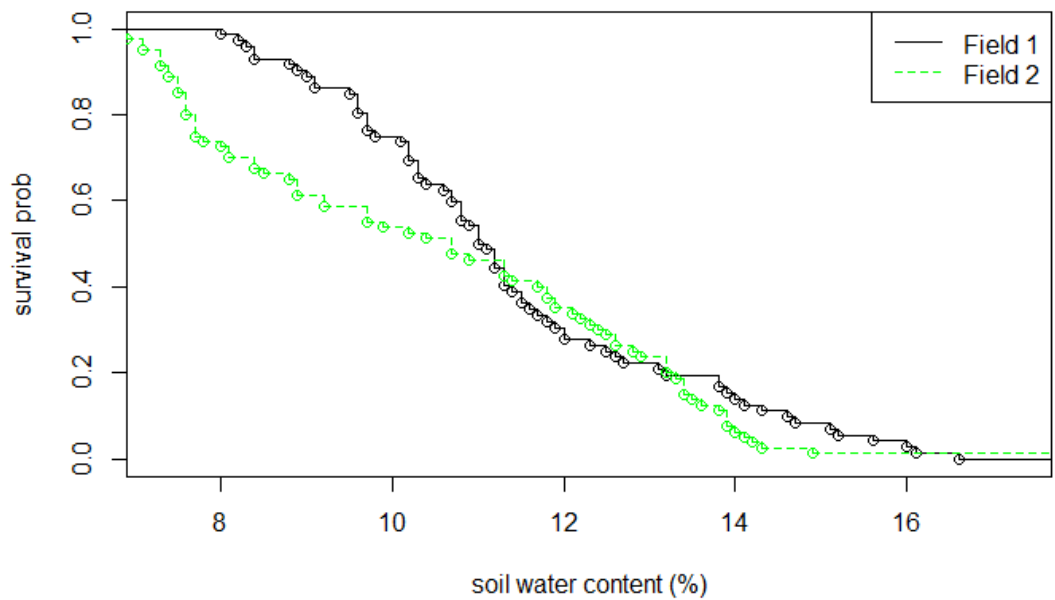

Figure 3: The empirical survival curves of soil water contents in two fields with Field 1 (solid line) and Field 2 (dashed line).

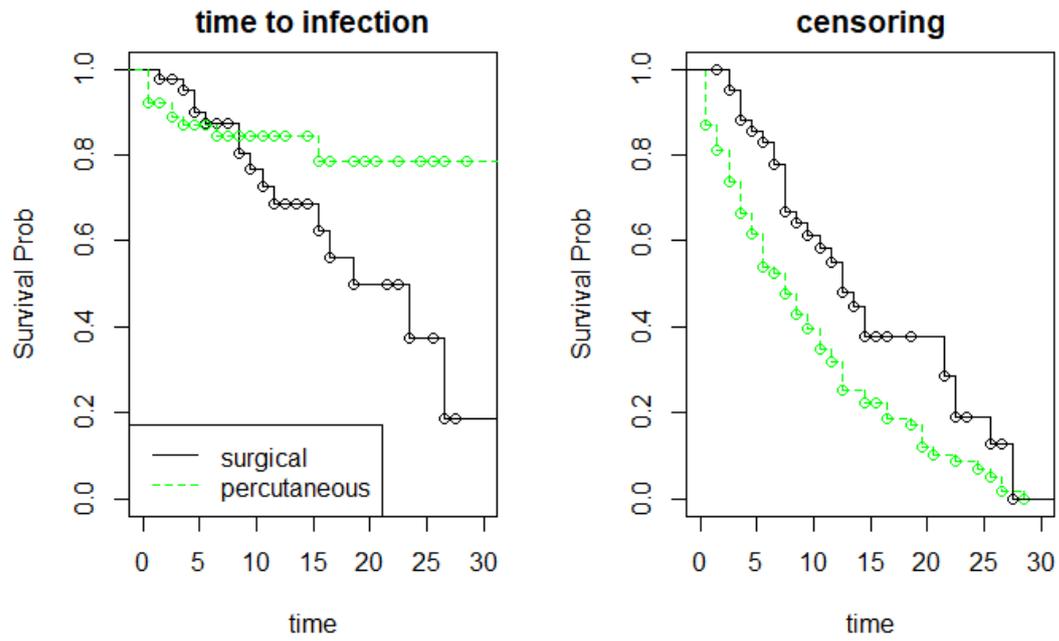

Figure 4: Left plot shows the KM curves of the infection times for patients utilizing percutaneous catheter placement (Group 0, dotted line) and surgical catheter placement (Group 1, solid line). Right plot shows the KM curves of the censoring distributions.

Table 1: Summary statistics of $\bar{\tau}_b$ and $\hat{\tau}_b$ based on $n = 400$ under the random grouping design. For the censored settings, censoring rates for Group 0 and Group 1 are (a) 0.50, 0.09; (b) 0.50, 0.33; (c) 0.5, 0.5; (d) 0.50, 0.67. "Avg. Bias" is the average bias shown in $\times 10^{-3}$ with the standard deviation in parenthesis and "C. Prob." is the coverage probability of 95% confidence intervals with the average length in parenthesis based on 2000 runs.

|  |  | Complete ($\bar{\tau}_b$) |  | Censored ($\hat{\tau}_b$) |  |
|---|---|---|---|---|---|
| Setting | Group Proportion | Avg. Bias (SD) | C. Prob. (Length) | Avg. Bias (SD) | C. Prob. (Length) |
| (a) $\kappa_0 = \kappa_1 = 1$ $\lambda_1 = 10$ $\lambda_0 = 1$ $\tau_b \approx -0.82$ | $p_1 = 0.4$ | -0.039 (0.029) | 0.940 (0.113) | 0.043 (0.029) | 0.952 (0.117) |
|  | $p_1 = 0.5$ | -0.199 (0.031) | 0.935 (0.119) | -0.420 (0.031) | 0.945 (0.123) |
|  | $p_1 = 0.7$ | 0.262 (0.038) | 0.929 (0.146) | -1.319 (0.039) | 0.932 (0.150) |
| (b) $\kappa_0 = \kappa_1 = 1$ $\lambda_1 = 2$ $\lambda_0 = 1$ $\tau_b \approx -0.33$ | $p_1 = 0.4$ | 0.730 (0.053) | 0.946 (0.210) | -0.456 (0.061) | 0.948 (0.240) |
|  | $p_1 = 0.5$ | 0.302 (0.053) | 0.948 (0.211) | -1.429 (0.062) | 0.948 (0.240) |
|  | $p_1 = 0.7$ | 0.737 (0.064) | 0.937 (0.240) | -1.709 (0.072) | 0.935 (0.271) |
| (c) $\kappa_0 = \kappa_1 = 1$ $\lambda_1 = 1$ $\lambda_0 = 1$ $\tau_b = 0$ | $p_1 = 0.4$ | 1.044 (0.059) | 0.945 (0.231) | -1.776 (0.073) | 0.940 (0.285) |
|  | $p_1 = 0.5$ | 0.419 (0.058) | 0.948 (0.227) | -2.209 (0.073) | 0.941 (0.279) |
|  | $p_1 = 0.7$ | 1.146 (0.065) | 0.941 (0.247) | -2.021 (0.081) | 0.936 (0.304) |
| (d) $\kappa_0 = \kappa_1 = 1$ $\lambda_1 = 0.5$ $\lambda_0 = 1$ $\tau_b \approx 0.33$ | $p_1 = 0.4$ | 1.164 (0.056) | 0.947 (0.220) | -3.927 (0.076) | 0.949 (0.295) |
|  | $p_1 = 0.5$ | 0.479 (0.054) | 0.947 (0.211) | -4.752 (0.074) | 0.945 (0.286) |
|  | $p_1 = 0.7$ | 1.220 (0.057) | 0.940 (0.219) | -5.782 (0.080) | 0.936 (0.302) |

Table 2: Summary Statistics of $\hat{\tau}_b^R$ and four versions of $\hat{\tau}_b^*$ with $\tau_b = 0\,(S_1 \neq S_0)$ and the true distributions being Weibull under $\xi_T > \xi_C$ and the random grouping design. Each cell contains the average and the standard deviation (in parenthesis) of $\hat{\tau}_b^*$ based on 2000 runs.

|  | $t^* = 0.5$ | $t^* = 0.8$ | $t^* = 1$ |
|---|---|---|---|
| $\tau_{b,2}^*$ | 0.2749 | 0.2186 | 0.1679 |
| $\hat{\tau}_b^*$: Weibull | 0.0012 (0.1340) | 0.0012 (0.1335) | 0.0123 (0.1385) |
| $\hat{\tau}_b^*$: Exponential | 0.4784 (0.1119) | 0.3491 (0.1177) | 0.2752 (0.1322) |
| $\hat{\tau}_b^*$: Log-normal | -0.0057 (0.1283) | -0.0043 (0.1307) | -0.0084 (0.1411) |
| $\hat{\tau}_b^*$: Logistic | 0.3089 (0.1431) | 0.2009 (0.1259) | 0.1649 (0.1283) |
| $\hat{\tau}_b^R$ | 0.2491 (0.1586) | 0.2491 (0.1586) | 0.2491 (0.1586) |

**Supporting Information for "Kendall's Tau for Two-sample Inference Problems"**

by Yi-Cheng Tai, Weijing Wang and Martin T. Wells

**Web Appendix A: Properties of $\bar{\tau}_b$**

*Web Appendix A1: Proof of Theorem 1*

Under the fixed grouping design, $N_\ell = n_\ell$ ($\ell = 0,1$) are fixed constants. The expression in equation (1) in the main text can be written as $U = \sum_{i=1}^{n_0}\sum_{j=1}^{n_1} h(T_j^{(1)}; T_i^{(0)})$, where the kernel function

$$h(T^{(1)}; T^{(0)}) = I(T^{(0)} < T^{(1)}) - I(T^{(0)} > T^{(1)})$$

satisfies $E\left[h^2(T^{(1)}; T^{(0)})\right] < \infty$ since it is bounded. The exact variance formula for $\bar{\tau}_b = U/(n_0 n_1)$ is given by

$$\frac{1}{n_0 n_1}\left\{(n_1 - 1)\sigma_{0,1}^2 + (n_0 - 1)\sigma_{1,0}^2 + \sigma_{1,1}^2\right\},$$

where each term equals

$$\begin{aligned}
\sigma_{0,1}^2 &= Cov\left\{h\left(T_1^{(1)}; T_1^{(0)}\right), h\left(T_2^{(1)}; T_1^{(0)}\right)\right\} \\
&= E\left\{h\left(T_1^{(1)}; T_1^{(0)}\right) h\left(T_2^{(1)}; T_1^{(0)}\right)\right\} - \tau_b^2 \\
&= E\left[E\left\{h\left(T_1^{(1)}; T_1^{(0)}\right) h\left(T_2^{(1)}; T_1^{(0)}\right) | T_1^{(0)}\right\}\right] - \tau_b^2 \\
&= E\left[E\left\{h\left(T_1^{(1)}; T_1^{(0)}\right) | T_1^{(0)}\right\} E\left\{h\left(T_2^{(1)}; T_1^{(0)}\right) | T_1^{(0)}\right\}\right] - \tau_b^2 \\
&= E\left[\left\{S_1\left(T^{(0)}\right) - F_1\left(T^{(0)}\right)\right\}^2\right] - \tau_b^2,
\end{aligned}$$

$$\sigma_{1,0}^2 = E\left[\left\{F_0\left(T^{(1)}\right) - S_0\left(T^{(1)}\right)\right\}^2\right] - \tau_b^2,$$

and

$$\sigma_{1,1}^2 = Cov\left\{h\left(T_1^{(1)}; T_1^{(0)}\right), h\left(T_1^{(1)}; T_1^{(0)}\right)\right\} = E\left\{h^2\left(T_1^{(1)}; T_1^{(0)}\right)\right\} - \tau_b^2 = 1 - \tau_b^2.$$

The variance of $\sqrt{n}(\bar{\tau}_b - \tau_b)$, denoted as $n\sigma^2_{\bar{\tau}_{b,F}}$, then becomes

$$\frac{n_0 + n_1}{n_0 n_1}\left\{(n_1-1)\sigma^2_{0,1} + (n_0-1)\sigma^2_{1,0} + \sigma^2_{1,1}\right\}$$

which in turn converges to $\frac{1}{p_0}\sigma^2_{0,1} + \frac{1}{p_1}\sigma^2_{1,0}$ as $n_0, n_1 \to \infty$.

*Web Appendix A2: Asymptotic Properties of $\sqrt{n}(\bar{\tau}_a - \tau_a)$*

**Lemma:** *The estimator $\bar{\tau}_a$ can be represented as a one-sample U-statistics of order 2 with the kernel function:*

$$\phi(\mathbf{T}_i, \mathbf{T}_j) = I\left\{(X_i - X_j)(T_i - T_j) > 0\right\} - I\left\{(X_i - X_j)(T_i - T_j) < 0\right\},$$

*where $\mathbf{T}_i = (X_i, T_i)$ with $E\{\phi^2(\mathbf{T}_i, \mathbf{T}_j)\} < \infty$ $(i \neq j)$. It follows that $\sqrt{n}(\bar{\tau}_a - \tau_a)$ converges to a mean-zero normal random variable with variance $n\sigma^2_{\bar{\tau}_a} = 4\sigma^2_1$, where*

$$\sigma^2_1 = Cov\{\phi(\mathbf{T}_1, \mathbf{T}_2), \phi(\mathbf{T}_1, \mathbf{T}_3)\}.$$

This result follows once derive the variance term. We can develop $\sigma^2_1$ as follows. By its definition,

$$\begin{aligned}
\sigma^2_1 &= E\{\phi(\mathbf{T}_1, \mathbf{T}_2)\phi(\mathbf{T}_1, \mathbf{T}_3)\} - \tau^2_a \\
&= E\left[\{E(\phi(\mathbf{T}_1, \mathbf{T}_2) | \mathbf{T}_1)\}^2\right] - \tau^2_a \\
&= E\left[\{E(\phi(\mathbf{T}_1, \mathbf{T}_2) | \mathbf{T}_1)\}^2\right] - (2p_0 p_1 \tau_b)^2, \quad (S1)
\end{aligned}$$

where

$$E(\phi(\mathbf{T}_1, \mathbf{T}_2) | \mathbf{T}_1) = p_1 I(X_1 = 0)\{S_1(T_1) - F_1(T_1)\} + p_0 I(X_1 = 1)\{F_0(T_1) - S_0(T_1)\}.$$

*Web Appendix A3: Proof of Theorem 2*

For the first component of the expression in (3) in the main text, it follows that

$$\sqrt{n}\left(\frac{1}{N_0 N_1} - \frac{1}{n^2 p_0 p_1}\right) U = \sqrt{n}\left(\frac{N_1}{n} - p_1\right)\left(\frac{p_1 - p_0}{p_0 p_1}\right)\tau_b + o_P(1),$$

which converges to a mean zero normal distribution with variance equal to $(p_0 p_1)^{-1}(p_1 - p_0)^2 \tau_b^2$. Since the second component of (3) is asymptotically equivalent to $\sqrt{n}(\bar{\tau}_a - \tau_a)/(2 p_0 p_1)$, it converges in distribution to a mean-zero normal variable with variance equal to $(p_0 p_1)^{-2} \sigma_1^2$, where $\sigma_1^2$ is given in (S1). The limiting covariance between the two terms in (3) can be derived as follows:

$$\begin{aligned}
&Cov\left[\left(\frac{p_1 - p_0}{p_0 p_1}\tau_b\right)\sqrt{n}\left(\frac{N_1}{n} - p_1\right), \frac{\sqrt{n}}{n^2 p_0 p_1}\sum_{1 \leq i < j \leq n} sign\{(X_i - X_j)(T_i - T_j)\}\right] \\
&= \tau_b \frac{p_1 - p_0}{n^2 p_0^2 p_1^2} Cov\left[\sum_{k=1}^n I(X_k = 1), \sum_{1 \leq i < j \leq n} sign\{(X_i - X_j)(T_i - T_j)\}\right] \\
&= \tau_b \frac{p_1 - p_0}{n^2 p_0^2 p_1^2} \sum_{1 \leq i < j \leq n} Cov\left[I(X_i = 1) + I(X_j = 1), sign\{(X_i - X_j)(T_i - T_j)\}\right] \\
&= \tau_b \frac{p_1 - p_0}{n^2 p_0^2 p_1^2} \sum_{1 \leq i < j \leq n} (\tau_a - 2 p_1 \tau_a) \\
&= \tau_b \frac{p_1 - p_0}{n^2 p_0^2 p_1^2} \binom{n}{2} \tau_a (1 - 2 p_1) \\
&= -\tau_b^2 \frac{(p_1 - p_0)^2}{p_0 p_1} + o(1).
\end{aligned}$$

Combining the previous results, it follows that $\sqrt{n}(\bar{\tau}_b - \tau_b)$ converges in distribution to a mean-zero normal variable with variance equal to

$$n\sigma_{\bar{\tau}_{b,R}}^2 = \frac{\sigma_1^2}{(p_0 p_1)^2} - \frac{(p_1 - p_0)^2}{p_0 p_1}\tau_b^2,$$

where $\sigma_1^2$ is given in (S1). The above analysis indicates that when $\tau_b = 0$ or $p_0 = p_1$, $\sqrt{n}\left(\frac{U}{N_0 N_1} - \frac{U}{n^2 p_0 p_1}\right)$ and $Cov\left[\sqrt{n}\left(\frac{U}{N_0 N_1} - \frac{U}{n^2 p_0 p_1}\right), \sqrt{n}\left(\frac{U}{n^2 p_0 p_1} - \tau_b\right)\right]$ both converge to zero in probability.

**Web Appendix B: Unified Expressions of the Proposed, Log Rank and Gehan Statistics**

*Web Appendix B1: Weighted Log Rank Statistics: Martingale Representations*

Define the counting process $N_\ell(t) = \sum_{i=1}^{n} I(Y_i \leq t, \delta_i = 1, X_i = \ell)$ and the at-risk process $R_\ell(t) = \sum_{i=1}^{n} I(Y_i \geq t, X_i = \ell)$ for Group $\ell$ ($\ell = 0, 1$). Let $N(t) = N_0(t) + N_1(t)$ and $R(t) = R_0(t) + R_1(t)$. Let $\lambda_\ell(t)$ and $\Lambda_\ell(t)$ be the hazard and cumulative hazard functions of $T \mid X = \ell$, respectively. Using the Nelson-Aalen approach we can estimate $\Lambda_\ell(t)$ by

$$\hat{\Lambda}_\ell(t) = \int_0^t I\{R_\ell(u) > 0\} \{R_\ell(u)\}^{-1} dN_\ell(u) \quad (\ell = 0, 1).$$

The WLR statistics can then be expressed as

$$U_*(t) = \int_0^t W_*(u) \left\{ dN_0(u) - \frac{R_0(u)}{R(u)} dN(u) \right\} = \int_0^t w_*(u) \left\{ d\hat{\Lambda}_0(u) - d\hat{\Lambda}_1(u) \right\}, \quad (S2)$$

where $w_*$ and $W_*(u)$ are weight functions such that $w_*(u) = W_*(u) \frac{R_0(u) R_1(u)}{R(u)}$. In this more general formulation the proposed estimator of $\tau_b$ can then be written as $\hat{\tau}_b = U_P(\infty) / (N_0 N_1)$, where $w_P(u) = \frac{R_0(u) R_1(u)}{\hat{G}_0(u) \hat{G}_1(u)}$. The LR and Gehan statistics, denoted as $U_L(\infty)$ and $U_G(\infty)$, have the weight functions $w_L(u) = \frac{R_0(u) R_1(u)}{R(u)}$ and $w_G(u) = R_0(u) R_1(u)$, respectively. The three WLR statistics $U_P(\infty)$, $U_G(\infty)$ and $U_L(\infty)$ are all valid for testing $H_0 : \Lambda_0(t) = \Lambda_1(t)$ or equivalently $H_0 : S_0(t) = S_1(t)$ for $t \leq \xi_Y$, where $\xi_Y = \sup\{t : \Pr(Y \geq t) > 0\}$.

From (S2), we can further write

$$U_*(\infty) = \left\{\int_0^\infty \frac{w_*(u)}{R_0(u)} dM_0(u) - \int_0^\infty \frac{w_*(u)}{R_1(u)} dM_1(u)\right\} + \int_0^\infty w_*(u)\{d\Lambda_0(u) - d\Lambda_1(u)\}$$
$$= U_*^{(1)}(\infty) + U_*^{(2)}(\infty),$$

where $M_0$ and $M_1$ are mean-zero martingales defined as

$$M_\ell(t) = N_\ell(t) - \int_0^t R_\ell(u) d\Lambda_\ell(u) \quad (\ell = 0, 1).$$

Since $U_*^{(1)}(\infty)$ is the difference of two mean-zero martingale integrals it is also a mean-zero martingale. Under $H_0 : \Lambda_0 = \Lambda_1$, it is easy to see that the second term $U_*^{(2)}(\infty) = 0$. When $H_0 : \Lambda_0 = \Lambda_1$ is false but $H_0^{tau} : \tau_b = 0$, $U_L^{(2)}(\infty)$ and $U_G^{(2)}(\infty)$ are usually not zero. This indicates that the LR and Gehan statistics are not suitable for testing $H_0^{tau} : \tau_b = 0$. However,

$n^{-2} \int_0^\infty w_P(u)\{d\Lambda_0(u) - d\Lambda_1(u)\}$ converges to

$$p_0 p_1 \left\{\int_0^\infty \Pr(T \geq u, X = 1) d\Pr(T \leq u, X = 0) - \int_0^\infty \Pr(T \geq u, X = 0) d\Pr(T \leq u, X = 1)\right\},$$

which is zero under $H_0^{tau} : \tau_b = 0$.

The following derivations show that the LR and Gehan statistics are strongly affected by the censoring distributions while the proposed statistic is not. The limits of the weight functions are given by

$$\lim_{n \to \infty} n^{-1} w_L(u) = \left[\{p_0 S_0(u-) G_0(u-)\}^{-1} + \{p_1 S_1(u-) G_1(u-)\}^{-1}\right]^{-1},$$

$$\lim_{n \to \infty} n^{-2} w_G(u) = p_0 p_1 S_0(u-) S_1(u-) G_0(u-) G_1(u-),$$

$$\lim_{n \to \infty} n^{-2} w_P(u) = p_0 p_1 S_0(u-) S_1(u-).$$

Using the limiting form above, notice that $n^{-2} U_P(t)$ converges to

$$p_0 p_1 \left\{ \int_0^t S_1(u-)dF_0(u) - \int_0^t S_0(u-)dF_1(u) \right\}.$$

Efron (1967) suggested to use $-\int_0^\infty \hat{S}_1(u-)d\hat{S}_0(u)$ for testing $H_0: S_0(t) = S_1(t)$ and this statistic seems to coincide with $U_P(\infty)$. Nevertheless, this article focuses on the role of $\hat{\tau}_b = U_P(\infty)/(N_0 N_1)$ as an estimator of $\tau_b$. In Web Appendix C, we will derive the properties of $\hat{\tau}_b$ under general conditions when $\tau_b \neq 0$ and $N_\ell$ is fixed or random. The results can be used to construct confidence intervals of $\tau_b$.

*Web Appendix B2: Linear Rank Statistics: Martingale Representations*

The LR and Gehan statistics have been compared under the context of linear rank statistics (Prentice 1978; Prentice and Marek, 1979). Each observation in the pooled sample gets a score such that the sum of total scores in the combined sample is zero. Then the chosen statistic is the sum of total scores assigned to all observations in one group. Here we re-express the results of the LR and Gehan statistics in the paper by Tarone and Ware (1977) using the counting process notations defined earlier and also present the formula of the proposed statistic. The score formula for the three statistics are given below:

$$V_i^L(t) = \int_0^t \frac{R_i(s)}{R(s)} dN(s) - \int_0^t dN_i(s),$$

$$V_i^G(t) = \int_0^t R_i(s) dN(s) - \int_0^t R(s) dN_i(s),$$

$$V_i^P(t) = \int_0^t \frac{R_i(s)}{\hat{G}_0(s)\hat{G}_1(s)} dN(s) - \int_0^t \frac{R(s)}{\hat{G}_0(s)\hat{G}_1(s)} dN_i(s),$$

where "L", "G" and "P" denote the LR, Gehan and proposed statistics, respectively. Given that $\sum_{i=1}^{n} V_i^*(\infty) = 0$ ( $* = "L, G, P"$ ), one can use $\sum_{i=1}^{n} X_i V_i^*(\infty)$ or $\sum_{i=1}^{n} (1-X_i) V_i^*(\infty)$ to compare the two groups. Compared with $V_i^L(\infty)$, we see that $V_i^G(\infty)$ down-weights an observation with larger failure time, while $V_i^P(\infty)$ removes the selection bias due to censoring by employing the technique of inverse probability weighting. Hence the proposed method is adaptive with respect to the censoring distribution.

*Web Appendix B3: Weighted Log Rank Statistics: U-Statistic Representations*

The LR, Gehan and proposed statistics can also be expressed in terms of the following U-statistics formula:

$$\sum_{i<j} \gamma_{ij}^* O_{ij} \, sign\{(X_i - X_j)(Y_i - Y_j)\}, \qquad (S3)$$

where $\gamma_{ij}^L = \left\{\sum_{k=1}^{n} I(Y_k \geq \tilde{Y}_{ij})\right\}^{-1}$, $\gamma_{ij}^G = 1$ and $\gamma_{ij}^P = \left\{\hat{G}_0(\tilde{Y}_{ij}) \hat{G}_1(\tilde{Y}_{ij})\right\}^{-1}$ correspond to the LR, Gehan and proposed statistics, respectively. In determining the order relationship of orderable pairs $(i, j)$, $\gamma_{ij}^G$ ignores the time information provided by $\tilde{Y}_{ij}$; while $\gamma_{ij}^L$ and $\gamma_{ij}^P$ both down-weight pairs with smaller $\tilde{Y}_{ij}$. The weight assigned by the LR statistic is the reciprocal of the number at risk at $\tilde{Y}_{ij}$. The weight assigned by the proposed statistic is the reciprocal of the joint probability of not being censored by time $\tilde{Y}_{ij}$, which is the weight suggested by inverse probability weight approaches.

# Web Appendix C: Properties of $\hat{\tau}_b$

*Web Appendix C1: Proof of Theorem 3*

Under the fixed grouping design, $N_\ell = n_\ell$ ($\ell = 0,1$). Consider the following decomposition:

$$\sqrt{n}(\hat{\tau}_b - \tau_b) = \sqrt{n}\left(\frac{\hat{U}}{n_0 n_1} - \frac{\breve{U}}{n_0 n_1}\right) + \sqrt{n}\left(\frac{\breve{U}}{n_0 n_1} - \tau_b\right), \tag{S4}$$

where

$$\breve{U} = \sum_{i=1}^{n_0}\sum_{j=1}^{n_1} \frac{\tilde{O}_{ij}\left\{I\left(Y_i^{(0)} < Y_j^{(1)}\right) - I\left(Y_i^{(0)} > Y_j^{(1)}\right)\right\}}{G_0\left(Y_i^{(0)} \wedge Y_j^{(1)}\right) G_1\left(Y_i^{(0)} \wedge Y_j^{(1)}\right)} = \sum_{i=1}^{n_0}\sum_{j=1}^{n_1} \psi\left(\mathbf{Y}_j^{(1)}; \mathbf{Y}_i^{(0)}\right). \tag{S5}$$

To simplify the notation of (S5), we denote $\tilde{Y}_{ij} = Y_i^{(0)} \wedge Y_j^{(1)}$ and $\psi_{ij} = \psi\left(\mathbf{Y}_j^{(1)}; \mathbf{Y}_i^{(0)}\right)$. For the first term in (S4), we can write

$$\sqrt{n}\left(\frac{\hat{U}}{n_0 n_1} - \frac{\breve{U}}{n_0 n_1}\right) = \frac{\sqrt{n}}{n_0 n_1}\sum_{i=1}^{n_0}\sum_{j=1}^{n_1}\left\{\frac{G_0(\tilde{Y}_{ij})G_1(\tilde{Y}_{ij})}{\hat{G}_0(\tilde{Y}_{ij})\hat{G}_1(\tilde{Y}_{ij})} - 1\right\}\psi_{ij}.$$

Note that

$$\frac{G_1(\tilde{Y}_{ij})G_0(\tilde{Y}_{ij})}{\hat{G}_1(\tilde{Y}_{ij})\hat{G}_0(\tilde{Y}_{ij})} - 1 = \frac{1}{n_0}\sum_{k=1}^{n_0}\int_0^{\tilde{Y}_{ij}} \frac{dM_{0,k}^C(u)}{\Pr(Y^{(0)} \geq u)} + \frac{1}{n_1}\sum_{k=1}^{n_1}\int_0^{\tilde{Y}_{ij}} \frac{dM_{1,k}^C(u)}{\Pr(Y^{(1)} \geq u)} + o_P\left(n^{-\frac{1}{2}}\right),$$

where $M_{\ell,k}^C(t) = I(Y_k^{(\ell)} \leq t, \delta_k^{(\ell)} = 0) - \int_0^t I\left(Y_k^{(\ell)} \geq u\right) d\Lambda_\ell^C(u)$ ($k = 1,...,n_\ell$) and $\Lambda_\ell^C(u)$ is the cumulative hazard function of $C^{(\ell)}$ ($\ell = 0,1$). Accordingly, the first component of (S4) can be expressed as

$$\sqrt{n}\left\{\frac{1}{n_0}\sum_{k=1}^{n_0}\int_0^\infty \frac{\eta(u)}{\Pr(Y^{(0)} \geq u)} dM_{0,k}^C(u) + \frac{1}{n_1}\sum_{k=1}^{n_1}\int_0^\infty \frac{\eta(u)}{\Pr(Y^{(1)} \geq u)} dM_{1,k}^C(u)\right\} + o_P(1),$$

where $\eta(u) = E\left\{\psi_{ij} I(\tilde{Y}_{ij} \geq u)\right\} = E\left\{\psi_{ij} I\left(Y_i^{(0)} \wedge Y_j^{(1)} \geq u\right)\right\}$. It follows that $\sqrt{n}\left(\frac{\hat{U}}{n_0 n_1} - \frac{\breve{U}}{n_0 n_1}\right)$ converges to a mean-zero normal random variable with variance equal to

$$\frac{1}{p_0}\int_0^\infty \frac{\eta^2(u)}{\Pr(Y^{(0)} \geq u)} d\Lambda_0^C(u) + \frac{1}{p_1}\int_0^\infty \frac{\eta^2(u)}{\Pr(Y^{(1)} \geq u)} d\Lambda_1^C(u). \qquad (S6).$$

For the second term of (S4), the classical two-sample U-statistics theory can be applied. It follows that

$$\Sigma_{0,1}^2 = Cov(\psi_{ij}, \psi_{ik}) = E(\psi_{ij}\psi_{ik}) - \tau_b^2 \quad (j \neq k),$$

$$\Sigma_{1,0}^2 = Cov(\psi_{ij}, \psi_{kj}) = E(\psi_{ij}\psi_{kj}) - \tau_b^2 \quad (i \neq k).$$

Hence $Var\left\{\sqrt{n}\left(\dfrac{\breve{U}}{n_0 n_1} - \tau_b\right)\right\}$ converges to

$$\frac{1}{p_0}\left\{\lim_{n_0,n_1\to\infty} \frac{1}{n_0 n_1^2} \sum_{i=1}^{n_0}\sum_{j,k}^{n_1} \psi_{ij}\psi_{ik} - \tau_b^2\right\} + \frac{1}{p_1}\left\{\lim_{n_0,n_1\to\infty} \frac{1}{n_1 n_0^2} \sum_{j=1}^{n_1}\sum_{i,k}^{n_0} \psi_{ij}\psi_{kj} - \tau_b^2\right\}. \qquad (S7)$$

The asymptotic covariance between $\sqrt{n}\left(\dfrac{\hat{U}}{n_0 n_1} - \dfrac{\breve{U}}{n_0 n_1}\right)$ and $\sqrt{n}\left(\dfrac{\breve{U}}{n_0 n_1} - \tau_b\right)$ can be derived

by simplifying the following expressions:

$$Cov\left[\sqrt{n}\left\{\frac{1}{n_0}\sum_{i=1}^{n_0}\int_0^\infty \frac{\eta(u)}{\Pr(Y^{(0)} \geq u)} dM_{0,i}^C(u) + \frac{1}{n_1}\sum_{j=1}^{n_1}\int_0^\infty \frac{\eta(u)}{\Pr(Y^{(1)} \geq u)} dM_{1,j}^C(u)\right\}, \frac{\sqrt{n}}{n_0 n_1}\sum_{i=1}^{n_0}\sum_{j=1}^{n_1}\psi_{ij}\right]$$

$$= \frac{n}{n_0 n_1}\sum_{i=1}^{n_0}\sum_{j=1}^{n_1} Cov\left\{\frac{1}{n_0}\int_0^\infty \frac{\eta(u)}{\Pr(Y^{(0)} \geq u)} dM_{0,i}^C(u) + \frac{1}{n_1}\int_0^\infty \frac{\eta(u)}{\Pr(Y^{(1)} \geq u)} dM_{1,j}^C(u), \psi_{ij}\right\}.$$

It follows that

$$E\left[\left\{\int_0^\infty \frac{\eta(u)}{\Pr(Y^{(0)} \geq u)} dM_{0,i}^C(u)\right\} \psi_{ij}\right]$$

$$= E\left\{\frac{\psi_{ij}\eta(Y_i^{(0)})}{\Pr(Y^{(0)} \geq Y_i^{(0)})} I(\delta_i^{(0)} = 0)\right\} - E\left\{\int_0^\infty \frac{\psi_{ij}\eta(u)}{\Pr(Y^{(0)} \geq u)} I(Y_i^{(0)} \geq u) d\Lambda_0^C(u)\right\}$$

$$= E\left\{\frac{\psi_{ij}\eta(C_i^{(0)})}{\Pr(Y^{(0)} \geq C_i^{(0)})} I(C_i^{(0)} < T_i^{(0)})\right\} - E\left\{\int_0^\infty \frac{\psi_{ij}\eta(u)}{\Pr(Y^{(0)} \geq u)} I(Y_i^{(0)} \geq u) d\Lambda_0^C(u)\right\}$$

$$= E\left[E\left\{\frac{\psi_{ij}\eta(C_i^{(0)})}{\Pr(Y^{(0)} \geq C_i^{(0)})} I(C_i^{(0)} < T_i^{(0)}) \middle| T_i^{(0)}, T_j^{(1)}, C_j^{(1)}\right\}\right] - E\left\{\int_0^\infty \frac{\psi_{ij}\eta(u)}{\Pr(Y^{(0)} \geq u)} I(Y_i^{(0)} \geq u) d\Lambda_0^C(u)\right\}$$

$$= E\left\{\int_0^\infty \frac{\psi_{ij}\eta(u)}{\Pr(Y^{(0)} \geq u)} I(T_i^{(0)} > u) G_0(u) d\Lambda_0^C(u)\right\} - E\left\{\int_0^\infty \frac{\psi_{ij}\eta(u)}{\Pr(Y^{(0)} \geq u)} I(Y_i^{(0)} \geq u) d\Lambda_0^C(u)\right\}$$

$$= E\left\{\int_0^\infty \frac{\psi_{ij}\eta(u)}{\Pr(Y^{(0)} \geq u)} I(Y_i^{(0)} > u) I(Y_j^{(1)} < u) d\Lambda_0^C(u)\right\} - E\left\{\int_0^\infty \frac{\psi_{ij}\eta(u)}{\Pr(Y^{(0)} \geq u)} I(Y_i^{(0)} \geq u) d\Lambda_0^C(u)\right\}$$

$$= -E\left\{\int_0^\infty \frac{\psi_{ij}\eta(u)}{\Pr(Y^{(0)} \geq u)} I(\tilde{Y}_{ij} \geq u) d\Lambda_0^C(u)\right\}.$$

Similarly,

$$E\left[\left\{\int_0^\infty \frac{\eta(u)}{\Pr(Y^{(1)} \geq u)} dM_{1,j}^C(u)\right\} \psi_{ij}\right] = -E\left\{\int_0^\infty \frac{\psi_{ij}\eta(u)}{\Pr(Y^{(1)} \geq u)} I(\tilde{Y}_{ij} \geq u) d\Lambda_1^C(u)\right\}.$$

We can then write

$$\frac{n}{n_0 n_1} \sum_{i=1}^{n_0} \sum_{j=1}^{n_1} \left[\frac{-1}{n_0} E\left\{\int_0^\infty \frac{\psi_{ij}\eta(u)}{\Pr(Y^{(0)} \geq u)} I(\tilde{Y}_{ij} \geq u) d\Lambda_0^C(u)\right\} + \frac{-1}{n_1} E\left\{\int_0^\infty \frac{\psi_{ij}\eta(u)}{\Pr(Y^{(1)} \geq u)} I(\tilde{Y}_{ij} \geq u) d\Lambda_1^C(u)\right\}\right]$$

$$= -n\left[\frac{1}{n_0} \int_0^\infty \frac{\eta^2(u)}{\Pr(Y^{(0)} \geq u)} d\Lambda_0^C(u) + \frac{1}{n_1} \int_0^\infty \frac{\eta^2(u)}{\Pr(Y^{(1)} \geq u)} d\Lambda_1^C(u)\right]$$

which converges to

$$-\frac{1}{p_0} \int_0^\infty \frac{\eta^2(u)}{\Pr(Y^{(0)} \geq u)} d\Lambda_0^C(u) - \frac{1}{p_1} \int_0^\infty \frac{\eta^2(u)}{\Pr(Y^{(1)} \geq u)} d\Lambda_1^C(u),$$

which is the negative of (S6).

Under the fixed grouping design in presence of censoring, the limiting variance of $\sqrt{n}(\hat{\tau}_b - \tau_b)$ can be analytically expressed as (S6) and (S7) such that

$$n\sigma^2_{\hat{\tau}_b,F} = \frac{1}{p_0}\left\{\lim_{n_0,n_1\to\infty}\frac{1}{n_0 n_1^2}\sum_{i=1}^{n_0}\sum_{j,k}^{n_1}\psi_{ij}\psi_{ik} - \tau_b^2\right\} + \frac{1}{p_1}\left\{\lim_{n_0,n_1\to\infty}\frac{1}{n_1 n_0^2}\sum_{j=1}^{n_1}\sum_{i,k}^{n_0}\psi_{ij}\psi_{kj} - \tau_b^2\right\}$$
$$- \frac{1}{p_0}\int_0^\infty \frac{\eta^2(u)}{\Pr(Y^{(0)}\geq u)}d\Lambda_0^C(u) - \frac{1}{p_1}\int_0^\infty \frac{\eta^2(u)}{\Pr(Y^{(1)}\geq u)}d\Lambda_1^C(u).$$

We can estimate $\sigma^2_{\hat{\tau}_b,F}$ by plugging in the estimate of each component. For example, $\eta(u)$ can be estimated by

$$\hat{\eta}(u) = \frac{1}{n_0 n_1}\sum_{i=1}^{n_0}\sum_{j=1}^{n_1}\left[\frac{\tilde{O}_{ij}\{I(Y_i^{(0)}<Y_j^{(1)}) - I(Y_i^{(0)}>Y_j^{(1)})\}}{\hat{G}_0(Y_i^{(0)}\wedge Y_j^{(1)})\hat{G}_1(Y_i^{(0)}\wedge Y_j^{(1)})}I(Y_i^{(0)}\wedge Y_j^{(1)}\geq u)\right],$$

and $\Pr(Y^{(\ell)}\geq u)$ can be estimated by $\sum_{k=1}^{n_\ell}I(Y_k^{(\ell)}\geq u)/n_\ell$ ($\ell = 0,1$).

*Web Appendix C2: Proof of Theorem 4*

Under the random grouping design, we adopt the pooled-sample expression of $\hat{U}$ in (5) of the main text. To simplify the notation, we write $\zeta_{ij} = \zeta(\mathbf{Y}_i, \mathbf{Y}_j)$. Similarly, $\breve{U}$ defined in (S5) can be re-expressed as

$$\breve{U} = \sum_{1\leq i<j\leq n}\frac{O_{ij}\text{sign}\{(X_i-X_j)(Y_i-Y_j)\}}{G_0(\tilde{Y}_{ij})G_1(\tilde{Y}_{ij})}, \tag{S8}$$

where $\tilde{Y}_{ij} = Y_i \wedge Y_j$. We can decompose $\sqrt{n}\left(\frac{\hat{U}}{N_0 N_1} - \tau_b\right)$ as follows:

$$\sqrt{n}\left(\frac{\hat{U}}{N_0 N_1} - \frac{\hat{U}}{n^2 p_0 p_1}\right) + \sqrt{n}\left(\frac{\hat{U}}{n^2 p_0 p_1} - \frac{\breve{U}}{n^2 p_0 p_1}\right) + \sqrt{n}\left(\frac{\breve{U}}{n^2 p_0 p_1} - \tau_b\right). \tag{S9}$$

The first term of (S9) reflects the effect of random grouping design, the second term is related to extra estimation of $G_0$ and $G_1$; and the third term is related to the distribution of the U-statistic estimator of $\tau_b$. For the first term in (S9), it follows that

$$\sqrt{n}\left(\frac{\hat{U}}{N_0 N_1} - \frac{\hat{U}}{n^2 p_0 p_1}\right) = \sqrt{n}\left(\frac{n^2 p_0 p_1}{N_0 N_1} - 1\right)\left(\frac{1}{n^2 p_0 p_1}\hat{U}\right)$$

$$= \sqrt{n}\left(\frac{N_1}{n} - p_1\right)\left\{\tau_b \frac{(p_1 - p_0)}{p_0 p_1}\right\} + o_P(1),$$

which converges in distribution to a mean-zero random variable with variance

$$\tau_b^2 \frac{(p_1 - p_0)^2}{p_0 p_1}. \tag{S10}$$

For the second term in (S9), we can write

$$\sqrt{n}\left(\frac{\hat{U}}{n^2 p_0 p_1} - \frac{\breve{U}}{n^2 p_0 p_1}\right) = \frac{\sqrt{n}}{n^2 p_0 p_1} \sum_{1 \leq i < j \leq n} \left\{\frac{G_1(\tilde{Y}_{ij}) G_0(\tilde{Y}_{ij})}{\hat{G}_1(\tilde{Y}_{ij}) \hat{G}_0(\tilde{Y}_{ij})} - 1\right\} \zeta_{ij}.$$

By the asymptotic martingale expressions of $\hat{G}_0$ and $\hat{G}_1$ presented earlier, the second term of (S9) can be written as

$$\sqrt{n}\left(\frac{\hat{U}}{n^2 p_0 p_1} - \frac{\breve{U}}{n^2 p_0 p_1}\right)$$

$$= \frac{\sqrt{n}}{n^2 p_0 p_1} \sum_{1 \leq i < j \leq n}\left\{\frac{1}{n}\sum_{k=1}^n \int_0^{\tilde{Y}_{ij}} \frac{dM_{0,k}^C(u)}{\Pr(Y \geq u, X = 0)} + \frac{1}{n}\sum_{k=1}^n \int_0^{\tilde{Y}_{ij}} \frac{dM_{1,k}^C(u)}{\Pr(Y \geq u, X = 1)} + o_P\left(n^{-\frac{1}{2}}\right)\right\}\zeta_{ij}$$

$$= \sqrt{n}\left\{\frac{1}{n}\sum_{k=1}^n \int_0^\infty \frac{(n-1)}{2np_0 p_1}\frac{\kappa(u)}{\Pr(Y \geq u, X = 0)}dM_{0,k}^C(u) + \frac{1}{n}\sum_{k=1}^n \int_0^\infty \frac{(n-1)}{2np_0 p_1}\frac{\kappa(u)}{\Pr(Y \geq u, X = 1)}dM_{1,k}^C(u)\right\}$$

$$+ o_P(1),$$

where $\kappa(u) = E\{\zeta_{ij} I(\tilde{Y}_{ij} \geq u)\}$, and it converges in distribution to a mean-zero normal random variable with variance equal to

$$\frac{1}{4 p_0^2 p_1^2}\left\{\int_0^\infty \frac{\kappa^2(u)}{\Pr(Y \geq u, X = 0)}d\Lambda_0^C(u) + \int_0^\infty \frac{\kappa^2(u)}{\Pr(Y \geq u, X = 1)}d\Lambda_1^C(u)\right\}. \tag{S11}$$

To analyze the third term of (S9), U-statistics theory is applicable. In particular, note that the quantity $\sqrt{n}\left(\dfrac{\breve{U}}{n^2 p_0 p_1} - \tau_b\right)$ converges to a mean-zero normal random variable with variance $\dfrac{\theta_1^2}{p_0^2 p_1^2}$, where

$$\theta_1^2 = \lim_{n\to\infty} \frac{1}{n^3} \sum_{i,j,k} \zeta_{ij}\zeta_{ik} - \tau_a^2 = \lim_{n\to\infty} \frac{1}{n^3} \sum_{i,j,k} \zeta_{ij}\zeta_{ik} - (2 p_0 p_1 \tau_b)^2. \tag{S12}$$

To obtain the limiting variance of $\sqrt{n}(\hat{\tau}_b - \tau_b)$, we still need to derive the limits of the following three covariance terms:

$$Cov\left\{\sqrt{n}\left(\frac{\hat{U}}{N_0 N_1} - \frac{\hat{U}}{n^2 p_0 p_1}\right), \sqrt{n}\left(\frac{\hat{U}}{n^2 p_0 p_1} - \frac{\breve{U}}{n^2 p_0 p_1}\right)\right\},$$

$$Cov\left\{\sqrt{n}\left(\frac{\hat{U}}{N_0 N_1} - \frac{\hat{U}}{n^2 p_0 p_1}\right), \sqrt{n}\left(\frac{\breve{U}}{n^2 p_0 p_1} - \tau_b\right)\right\},$$

$$Cov\left\{\sqrt{n}\left(\frac{\hat{U}}{n^2 p_0 p_1} - \frac{\breve{U}}{n^2 p_0 p_1}\right), \sqrt{n}\left(\frac{\breve{U}}{n^2 p_0 p_1} - \tau_b\right)\right\}.$$

Asymptotically $Cov\left\{\sqrt{n}\left(\dfrac{\hat{U}}{N_0 N_1} - \dfrac{\hat{U}}{n^2 p_0 p_1}\right), \sqrt{n}\left(\dfrac{\hat{U}}{n^2 p_0 p_1} - \dfrac{\breve{U}}{n^2 p_0 p_1}\right)\right\}$ equals

$$Cov\left[\tau_b \frac{(p_1 - p_0)}{p_0 p_1} \sqrt{n}\left(\frac{N_1}{n} - p_1\right), \sqrt{n}\left\{\frac{1}{n}\sum_{k=1}^n \int_0^\infty \frac{1}{2 p_0 p_1} \frac{\kappa(u)}{\Pr(Y \geq u, X = 0)} dM_{0,k}^C(u) \right.\right.$$

$$\left.\left. + \frac{1}{n}\sum_{k=1}^n \int_0^\infty \frac{1}{2 p_0 p_1} \frac{\kappa(u)}{\Pr(Y \geq u, X = 1)} dM_{1,k}^C(u)\right\}\right].$$

Notice that

$$Cov\left\{N_1, \sum_{k=1}^{n} \int_0^{\infty} \frac{1}{2p_0 p_1} \frac{\kappa(u)}{\Pr(Y \geq u, X = 0)} dM_{0,k}^C(u)\right\}$$

$$= \sum_{i=1}^{n} \sum_{k=1}^{n} E\left\{I(X_i = 1) \int_0^{\infty} \frac{1}{2p_0 p_1} \frac{\kappa(u)}{\Pr(Y \geq u, X = 0)} dM_{0,k}^C(u)\right\}$$

$$= \sum_{i=1}^{n} \sum_{k=1}^{n} E\left\{I(X_i = 1) \left[\frac{1}{2p_0 p_1} \frac{\kappa(Y_i)}{\Pr(Y \geq Y_i, X = 0)} I(X_i = 0, Y_i = u, \delta_i = 0)\right.\right.$$

$$\left.\left. - \int_0^{\infty} \frac{1}{2p_0 p_1} \frac{\kappa(u)}{\Pr(Y \geq u, X = 0)} I(Y_i \geq u, X_i = 0) d\Lambda_0^C(u)\right]\right\}$$

$$= 0.$$

Similarly,

$$Cov\left\{N_1, \sum_{k=1}^{n} \int_0^{\infty} \frac{1}{2p_0 p_1} \frac{\kappa(u)}{\Pr(Y \geq u, X = 1)} dM_{1,k}^C(u)\right\} = 0.$$

Hence, the covariance between the first two terms in (S9) converges to zero. The second covariance term $Cov\left\{\sqrt{n}\left(\frac{\hat{U}}{N_0 N_1} - \frac{\hat{U}}{n^2 p_0 p_1}\right), \sqrt{n}\left(\frac{\breve{U}}{n^2 p_0 p_1} - \tau_b\right)\right\}$ is asymptotically equal to

$$Cov\left[\sqrt{n}\left(\frac{N_1}{n} - p_1\right)\left\{\tau_b \frac{(p_1 - p_0)}{p_0 p_1}\right\}, \sqrt{n} \frac{\breve{U}}{n^2 p_0 p_1}\right],$$

which can be written as

$$\tau_b \frac{(p_1 - p_0)}{n^2 p_0^2 p_1^2} Cov\left\{\sum_{k=1}^{n} I(X_k = 1), \sum_{1 \leq i < j \leq n} \zeta_{ij}\right\}$$

$$= \tau_b \frac{(p_1 - p_0)}{n^2 p_0^2 p_1^2} \sum_{1 \leq i < j \leq n} Cov\left\{I(X_i = 1) + I(X_j = 1), \zeta_{ij}\right\}$$

$$= \tau_b \frac{(p_1 - p_0)}{n^2 p_0^2 p_1^2} \binom{n}{2} (1 - 2p_1) \tau_a$$

$$= -\tau_b^2 \frac{(p_1 - p_0)^2}{p_0 p_1},$$

which is the negative of (S10). Hence the covariance between the second and third terms in (S9) is asymptotically equal to

$$\mathrm{Cov}\left[\sqrt{n}\left\{\frac{1}{n}\sum_{k=1}^{n}\int_{0}^{\infty}\frac{1}{2p_0p_1}\frac{\kappa(u)}{\Pr(Y\geq u, X=0)}dM_{0,k}^{C}(u)\right.\right.$$
$$\left.\left.+\frac{1}{n}\sum_{k=1}^{n}\int_{0}^{\infty}\frac{1}{2p_0p_1}\frac{\kappa(u)}{\Pr(Y\geq u, X=1)}dM_{1,k}^{C}(u)\right\}, \frac{\sqrt{n}}{n^2 p_0 p_1}\sum_{1\leq i<j\leq n}\zeta_{ij}\right]$$
$$=\frac{1}{2n^2 p_0^2 p_1^2}\sum_{1\leq i<j\leq n}\mathrm{Cov}\left\{\sum_{k=1}^{n}\int_{0}^{\infty}\frac{\kappa(u)}{\Pr(Y\geq u, X=0)}dM_{0,k}^{C}(u)+\sum_{k=1}^{n}\int_{0}^{\infty}\frac{\kappa(u)}{\Pr(Y\geq u, X=1)}dM_{1,k}^{C}(u),\zeta_{ij}\right\}$$
$$=\frac{1}{2n^2 p_0^2 p_1^2}\sum_{1\leq i<j\leq n}\mathrm{Cov}\left[\int_{0}^{\infty}\frac{\kappa(u)}{\Pr(Y\geq u, X=0)}\{dM_{0,i}^{C}(u)+dM_{0,j}^{C}(u)\}\right.$$
$$\left.+\int_{0}^{\infty}\frac{\kappa(u)}{\Pr(Y\geq u, X=1)}\{dM_{1,i}^{C}(u)+dM_{1,j}^{C}(u)\},\zeta_{ij}\right].$$

It follows that

$$\mathrm{Cov}\left\{\int_{0}^{\infty}\frac{\kappa(u)}{\Pr(Y\geq u, X=0)}dM_{0,i}^{C}(u),\zeta_{ij}\right\}$$
$$=E\left\{\frac{\kappa(u)\zeta_{ij}}{\Pr(Y\geq u, X=0)}I(X_i=0, Y_i=u, \delta_i=0)\right\}-E\left\{\int_{0}^{\infty}\frac{\kappa(u)\zeta_{ij}}{\Pr(Y\geq u, X=0)}I(Y_i\geq u, X_i=0)d\Lambda_0^C(u)\right\}.$$

Note that

$$E\left\{\frac{\kappa(u)\zeta_{ij}}{\Pr(Y\geq u, X=0)}I(X_i=0, Y_i=u, \delta_i=0)\right\}=E\left\{\frac{\kappa(C_i)\zeta_{ij}}{\Pr(Y\geq C_i, X=0)}I(X_i=0, T_i>C_i)\right\},$$

which can be written as

$$E\left[E\left\{\frac{\kappa(C_i)\zeta_{ij}}{\Pr(Y\geq C_i, X=0)}I(X_i=0, T_i>C_i)\Big|X_i, X_j, T_i, T_j, C_j\right\}\right]$$
$$=E\left\{\int_{0}^{\infty}\frac{\kappa(u)\zeta_{ij}}{\Pr(Y\geq u, X=0)}I(X_i=0, T_i>u)G_0(u)d\Lambda_0^C(u)\right\}$$
$$=E\left\{\int_{0}^{\infty}\frac{\kappa(u)\zeta_{ij}}{\Pr(Y\geq u, X=0)}I(X_i=0, Y_i>u)I(X_j=1, Y_j<u)d\Lambda_0^C(u)\right\}.$$

Furthermore,

$$\mathrm{Cov}\left\{\int_{0}^{\infty}\frac{\kappa(u)}{\Pr(Y\geq u, X=0)}dM_{0,i}^{C}(u),\zeta_{ij}\right\}=-E\left\{\int_{0}^{\infty}\frac{\kappa(u)\zeta_{ij}I(\tilde{Y}_{ij}\geq u)}{\Pr(Y\geq u, X=0)}I(X_i=0, X_j=1)d\Lambda_0^C(u)\right\}.$$

Following similar arguments, we obtain

$$Cov\left\{\int_0^\infty \frac{\kappa(u)}{\Pr(Y \geq u, X = 0)} dM_{0,j}^C(u), \zeta_{ij}\right\} = -E\left\{\int_0^\infty \frac{\kappa(u)\zeta_{ij} I(\tilde{Y}_{ij} \geq u)}{\Pr(Y \geq u, X = 0)} I(X_i = 1, X_j = 0) d\Lambda_0^C(u)\right\},$$

$$Cov\left\{\int_0^\infty \frac{\kappa(u)}{\Pr(Y \geq u, X = 1)} dM_{0,i}^C(u), \zeta_{ij}\right\} = -E\left\{\int_0^\infty \frac{\kappa(u)\zeta_{ij} I(\tilde{Y}_{ij} \geq u)}{\Pr(Y \geq u, X = 1)} I(X_i = 1, X_j = 0) d\Lambda_1^C(u)\right\},$$

$$Cov\left\{\int_0^\infty \frac{\kappa(u)}{\Pr(Y \geq u, X = 1)} dM_{1,j}^C(u), \zeta_{ij}\right\} = -E\left\{\int_0^\infty \frac{\kappa(u)\zeta_{ij} I(\tilde{Y}_{ij} \geq u)}{\Pr(Y \geq u, X = 1)} I(X_i = 0, X_j = 1) d\Lambda_1^C(u)\right\}.$$

The limit of the covariance between the second and third terms of (S9) can be written as

$$-\frac{1}{4p_0^2 p_1^2}\left[\int_0^\infty \frac{\kappa^2(u)}{\Pr(Y \geq u, X = 0)} d\Lambda_0^C(u) + \int_0^\infty \frac{\kappa^2(u)}{\Pr(Y \geq u, X = 1)} d\Lambda_1^C(u)\right],$$

which is the negative of (S11).

Combining the results in (S10), (S11), and $\theta_1^2$ in (S12), the limiting variance of $\sqrt{n}(\hat{\tau}_b - \tau_b)$ under the random grouping design, denoted as $n\sigma_{\hat{\tau}_{b,R}}^2$, can be expressed analytically as

$$\frac{\theta_1^2}{p_0^2 p_1^2} - \frac{1}{4p_0^2 p_1^2}\left\{\int_0^\infty \frac{\kappa^2(u)}{\Pr(Y \geq u, X = 0)} d\Lambda_0^C(u) + \int_0^\infty \frac{\kappa^2(u)}{\Pr(Y \geq u, X = 1)} d\Lambda_1^C(u)\right\} - \tau_b^2 \frac{(p_1 - p_0)^2}{p_0 p_1}.$$

*Web Appendix C3: Asymptotic Variances under $H_0^{tau}: \tau_b = 0$*

First consider the case of fixed grouping design under $\tau_b = 0$ which also includes $S_0 = S_1$ as a special case. Based on Web Appendix B1, $\breve{U}$ in (S4) under $\tau_b = 0$ can be written as

$$\breve{U} = \int_0^\infty \frac{R_0(u)R_1(u)}{G_0(u)G_1(u)}\{R_0(u)\}^{-1} dM_0(u) - \int_0^\infty \frac{R_0(u)R_1(u)}{G_0(u)G_1(u)}\{R_1(u)\}^{-1} dM_1(u). \quad (S13)$$

Accordingly, the asymptotic variance of $\sqrt{n}\left(\dfrac{\breve{U}}{n_0 n_1} - \tau_b\right)$ is given by

$$\frac{1}{p_0}\int_0^\infty \left\{\frac{\Pr(Y^{(1)} \geq u)}{G_0(u)G_1(u)}\right\}^2 \Pr(Y^{(0)} \geq u)\,d\Lambda_0(u) + \frac{1}{p_1}\int_0^\infty \left\{\frac{\Pr(Y^{(0)} \geq u)}{G_0(u)G_1(u)}\right\}^2 \Pr(Y^{(1)} \geq u)\,d\Lambda_1(u)$$

which is used to replace (S7) in the formula of $\sigma^2_{\hat{\tau}_{b,F}}$ under $\tau_b = 0$.

Next consider the case of random grouping design under $\tau_b = 0$. In (S9), $\breve{U}$ is re-expressed as (S13) and the asymptotic variance of $\sqrt{n}\left(\dfrac{\breve{U}}{n^2 p_0 p_1} - \tau_b\right)$ is

$$\frac{1}{p_0^2 p_1^2} \times \int_0^\infty \left\{\frac{\Pr(Y \geq u, X = 1)}{G_0(u)G_1(u)}\right\}^2 \Pr(Y \geq u, X = 0)\,d\Lambda_0(u)$$

$$+ \frac{1}{p_0^2 p_1^2} \times \int_0^\infty \left\{\frac{\Pr(Y \geq u, X = 0)}{G_0(u)G_1(u)}\right\}^2 \Pr(Y \geq u, X = 1)\,d\Lambda_1(u),$$

which is used to replace $\dfrac{\theta_1^2}{p_0^2 p_1^2}$ in (S12) under $\tau_b = 0$ which appears in $n\sigma^2_{\hat{\tau}_{b,R}}$.

We claim that, when $\tau_b = 0$, the asymptotic variances of $\sqrt{n}(\hat{\tau}_b - \tau_b)$ under the fixed and random grouping designs are equal. It is easy to see that $\Pr(Y \geq u, X = \ell) = p_\ell \Pr(Y^{(\ell)} \geq u)$ for $\ell = 0,1$. Also note that $\kappa(u) = (2p_0 p_1)\eta(u)$, where $\eta(u) = \{\psi_{ij} I(Y_i^{(0)} \wedge Y_j^{(1)} \geq u)\}$ and $\kappa(u) = E\{\zeta_{ij} I(Y_i \wedge Y_j \geq u)\}$. Then we can show that $n\sigma^2_{\hat{\tau}_{b,F}}$ and $n\sigma^2_{\hat{\tau}_{b,R}}$ are equal under $\tau_b = 0$. This implies that the proposed test statistics $(\hat{\tau}_b - \tau_b)/\sigma_{\hat{\tau}_{b,F}}$ and $(\hat{\tau}_b - \tau_b)/\sigma_{\hat{\tau}_{b,R}}$ are asymptotically equivalent under $H_0^{tau}: \tau_b = 0$ and $H_0: S_0 = S_1$.

**Web Appendix D: Additional Simulation Results**

*Web Appendix D1: Performances of $\bar{\tau}_b$ and $\hat{\tau}_b$ with various sample sizes under two designs*

First we investigate the performances of $\bar{\tau}_b$, $\hat{\tau}_b$, $\hat{\sigma}_{\bar{\tau}_b}$ and $\hat{\sigma}_{\hat{\tau}_b}$ based on different sample sizes ($n = 100, 200$ and $400$) under both grouping designs. Under the fixed design, we generate $T_k^{(\ell)}$ and $C_k^{(\ell)}$ from independent exponential distributions for $k = 1,...,n_\ell$, where $n_\ell = np_\ell$ ($\ell = 0,1$) are pre-specified. Web Tables 1, 3 and 5 can be used to evaluate the performances of $\bar{\tau}_b$, $\hat{\tau}_b$, $\hat{\sigma}_{\bar{\tau}_{b,F}}$ and $\hat{\sigma}_{\hat{\tau}_{b,F}}$ under various configurations. Each table presents the average bias and standard deviation (in parenthesis) of $\bar{\tau}_b$ and $\hat{\tau}_b$; and the coverage probability and the length (in parenthesis) of 95% confidence intervals based on 2000 simulation runs. Web Tables 2, and 4 present analogous results with $n = 100$ and 200, respectively, based on the random grouping design. Notice that the length of the confidence interval becomes wider when the censoring rate increases. Given the same level of $p_1$, the length is widest when $\tau_b = 0$. The coverage probabilities are close to the nominal level 0.95 in most cases. In Web Table 1, we see that the coverage probability falls below 0.9 under the setting with $\tau_b \approx -0.82$, $p_1 = 0.7$ and $n = 100$ for both complete and censored data. Since the average biases of $\bar{\tau}_b$ and $\hat{\tau}_b$ are still reasonable, this implies that the normal approximation does not work well in this case with $n_1 \approx 70$ and $n_2 \approx 30$. Nevertheless, the normal approximation improves as $n = 200$ and 400. In Web Table 6, the results with $n = 400$ based on unequal censoring distributions are presented, where $C \mid X = \ell \sim Exp(\lambda_\ell^C)$, $\lambda_0^C = 1$ and $\lambda_0^C = 0.5$. The patterns are similar to the cases with equal censoring distributions.

Web Table 7 provides the detailed information of Figure 1 of the main paper. The results show that, when the PH assumption is valid, the power of the proposed $\hat{\tau}_b$ statistic is between those of the LR and the Gehan statistics. Web Table 8 gives the detailed information of Figure 2 in the main paper. The results indicate that, when the PH assumption does not hold, the distribution of the LR test statistic is much affected by the censoring rates. On the other hand, the distribution of the proposed $\hat{\tau}_b$ statistic is quite robust. Although the distributions of the Gehan statistic appears relatively robust under the current scenarios, it is because it assigns higher weight to small failure points which are less likely to be censored. Recall that the Gehan and the proposed $\hat{\tau}_b$ statistics both reduce to the WMW statistic when the censoring rates approach to zero. Since $\tau_b = 0$ under this setting, we can conclude that the Gehan statistic is still affected by the censoring distributions since the histogram is not centered around zero.

*Appendix D2: Modification for delayed treatment effect*

Xu et. al (2017) proposed a modified WLR test under the standard delayed treatment effect scenario such that $\lambda_0(t) = \lambda_1(t)$ for $t \leq t_0$ and $\lambda_0(t) = \rho \lambda_1(t)$ for $t > t_0$, where $t_0$ denotes the hazard changing point for the treatment group. We consider the more general scenario that $\lambda_0(t) \neq \lambda_1(t)$ for $t > t_0$ and modify the proposed methodology as follows. Specifically, we can use

$$U^{(D)} = \sum_{i<j} \frac{O_{ij} I(\tilde{Y}_{ij} > t_0) \, sign\{(X_i - X_j)(Y_i - Y_j)\}}{\hat{G}_0(\tilde{Y}_{ij}) \hat{G}_1(\tilde{Y}_{ij})},$$

which measures the concordant and discordant relationship after the changing point $t_0$. Analytic asymptotic variance formula of $U^{(D)}$ can be derived using the same techniques introduced earlier. We can further normalize $U^{(D)}$ to obtain an interpretable association measure located in $[-1,1]$ as follows:

$$\hat{\tau}_b^{(D)} = \left\{ \sum_{i<j} \frac{O_{ij} I(\tilde{Y}_{ij} > t_0)}{\hat{G}_0(\tilde{Y}_{ij}) \hat{G}_1(\tilde{Y}_{ij})} \right\}^{-1} U^{(D)}.$$

We run simulations to evaluate the performances of the proposed unmodified and modified statistics $\hat{\tau}_b$ and $U^{(D)}$ respectively under the setting by Xu et. al (2017). We generate $X \sim Bernoulli(p_1 = 0.5)$, $T | X = 0$ from $Exp(\lambda_0)$ and $T | X = 1$ as follows. First, generate $A \sim Exp(\lambda_0)$. Then if $A \leq t_0$, set $T = A$; if $A > t_0$, generate $B \sim Exp(\rho^{-1}\lambda_0)$ and set $T = t_0 + B$. We set $\lambda_0 = 1$, $t_0 = 1$ and $\rho = 0.1, 0.2, ..., 1$. For the censoring distributions, $C | X = 0$ and $C | X = 1$ both follow $Exp(1)$. The censoring rate $\Pr(\delta = 0 | X = \ell)$ under selected values of $\rho$ is between 0.5 and 0.55 for both $\ell = 0, 1$. Based on $n = 400$, we present the power curves of the test statistics $\hat{\tau}_b$ and $U^{(D)}$ together under selected values of $\rho$. For convenience, the variance of $U^{(D)}$ is estimated based on 500 bootstrapped samples. Web Figure 2 shows that the power of $U^{(D)}$ increases as $\rho$ departs more from 1. Although we don't know how $U^{(D)}$ is compared with the modified WLR test proposed by Xu et al. (2017), we believe that it also possesses the robustness property as $\hat{\tau}_b$.

## Web Appendix E: Additional Data Analysis

*Web Appendix E1: Gender-Age comparison for Covid-19 cases in Taiwan*

We apply the proposed methodology to analyze the age distributions of female and male Covid-19 infected cases in Taiwan from 2020/01/22 to 2020/12/31. Compared to the rest of the world, the Covid-19 pandemic had a much smaller impact in Taiwan during 2020. The dataset contains 366 males (Group 1) and 414 females (Group 0), most of whom were non-domestic and only 8 cases were infected locally. Due to the Personal Information Protection Act (Ministry of Justice, Taiwan 2012) the information of exact ages is not available and only 5-year age intervals are released. Therefore, we impute the age by generating a uniform random variable from a given age interval below 70 or $70 + Exp(0.5)$ for the age interval older than 70. We ran the imputation procedure many times and found that the results were similar so that we present the analysis based on one set of imputed data.

The plots in Web Figure 3 display the two gender survival curves of confirmed cases in 2020 and the curves in the three-time phases. The summary statistics are presented in Web Table 8A. The median ages for male and female cases are 33.62 and 30.91, respectively. The formula of $\hat{\sigma}_{\bar{\tau}_b, R}$ is adopted to construct confidence intervals. Except for Phase 2, the positive sign of $\bar{\tau}_b$ shows that male cases tend to be slightly older than female cases. Since most cases were imported and could be traced, we can compare the demographic patterns in the three phases and find out the reasons. In Phase 1 (1/22~4/17/2020), there are 184 infected males and 211 infected females with median ages 33.95 and 30.93 respectively. Many infected cases were Taiwanese businessmen and their relatives from China, Taiwanese who studied, worked or traveled overseas. The analysis implies that the male cases are slightly older than female cases but the difference is not significant. In Phase 2 (4/18~6/21/2020), there are 38 infected males and 13 infected females with median ages 25.22 and 33.53 respectively. Note that 29 cases were infected on a Taiwanese navy ship and most of them were male young soldiers. Although males are obviously younger, the difference is not significant due to the small sample sizes. In Phase 3 (6/22~12/31/2020), there are 144 infected males and 190 infected females with median ages 36.59 and 30.85 respectively. A large proportion of imported cases were female migrant care-workers due to the high demand. The hypothesis $H_0^{tau}: \tau_b = 0$ is rejected with the p-value equal to 0.0002. The estimated value $\bar{\tau}_b = 0.257$ indicate that the confirmed females are younger than males significantly.

In Web Table 9B, we present the original release data in the whole year of 2020 and the p-value based on the Chi-squared test is 0.0456 which indicates that gender and age are correlated. The Chi-squared test is also applied to test the gender and age relationship for Phase 1 and the p-value is 0.1404. The results based on the Chi-squared test are consistent with our analysis. However, our approach provides additional useful information about the strength and direction of the association. Furthermore, Chi-squared tests are not suitable if some cell counts are too small as in Phases 2 and 3.

*Web Appendix E2: Comparison of recurrence time for two treatments of bladder cancer*

The dataset is obtained from Pagano and Gauvreau (2000: exercise 9, page 512). There were 86 patients who received surgery to remove tumors. After the surgery, 48 patients were assigned to the placebo treatment (Group 0) and 38 patients were assigned to chemotherapy (Group 1). The variable under comparison is the time to first recurrence of tumor. In this dataset, 53% patients in the chemotherapy group and 40% in the placebo grouped did not experience recurrences by the end of the study. The overall censoring rate is 45% and the largest observed time of $Y$ is $y_{(n)} = 59$. From the two Kaplan-Meier curves in Web Figure 4, the two groups overlap at the beginning but differ in later time. The censoring rates are $\Pr(\delta = 0) \approx 0.45$ and $\Pr(\delta = 0 | X = \ell) \approx 0.4$ and $0.53$ for $\ell = 0, 1$, respectively. We obtain $\hat{\tau}_b = 0.145$ and $\hat{\tau}_b^R = 0.181$, which indicate that patients receiving chemotherapy tend to experience tumor recurrence at later time. Web Tables 10A and 10B list the values of $\hat{\tau}_b^*$ and the resulting bootstrapped confidence intervals with the tail imputed by four parametric models after selected values of $t^*$. The bootstrap confidence intervals are constructed based on 2000 runs. Since all the intervals contain zero, we can conclude that the two treatments have no significant difference. From Web Tables 10C, the three tests (i.e. LR, Gehan and proposed statistics) give the same conclusion of accepting $H_0 : S_0 = S_1$.

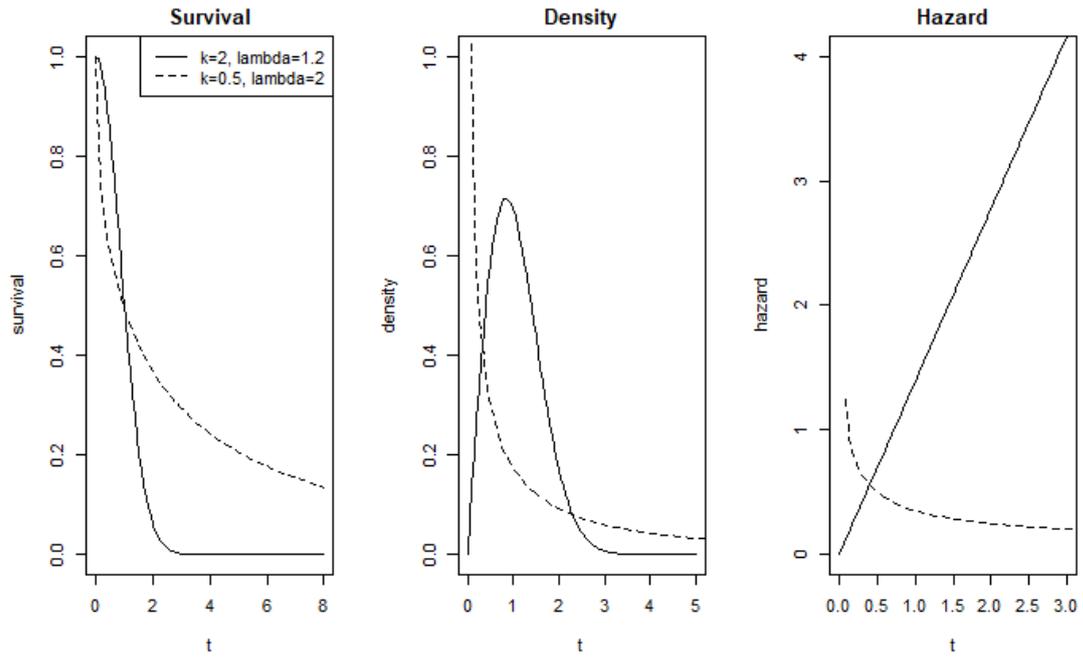

Web Figure 1: Survival (left), density (middle) and hazard (right) functions for two groups ("Solid": Group 0 and "Dashed": Group 1).

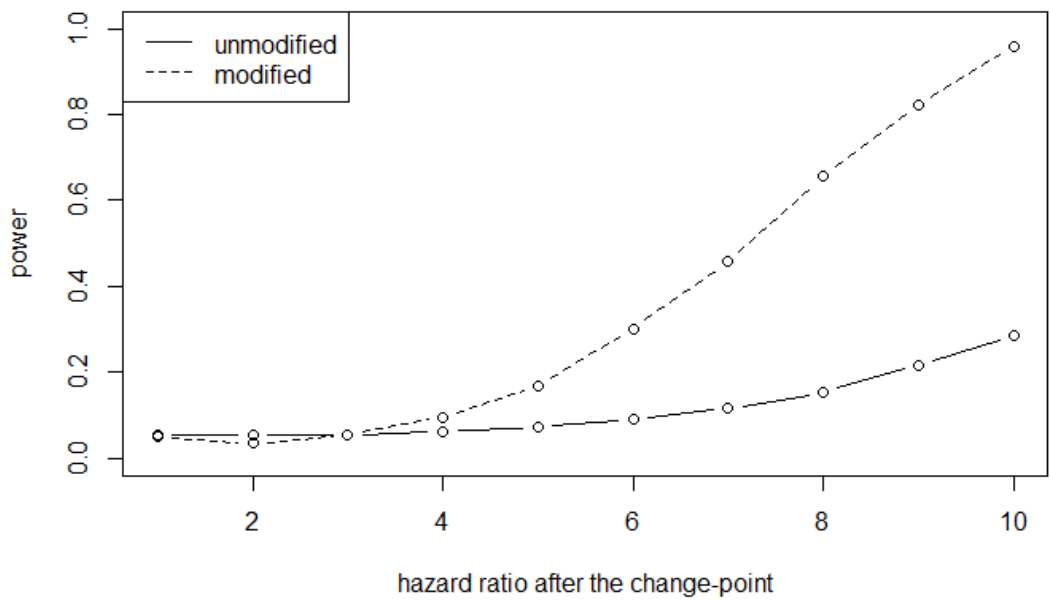

.

Web Figure 2: Power functions of $\hat{\tau}_b$ and $U^{(D)}$ under delayed treatment effect.

("Solid": un-modified statistic $\hat{\tau}_b$ and "Dashed": modified statistic $U^{(D)}$)

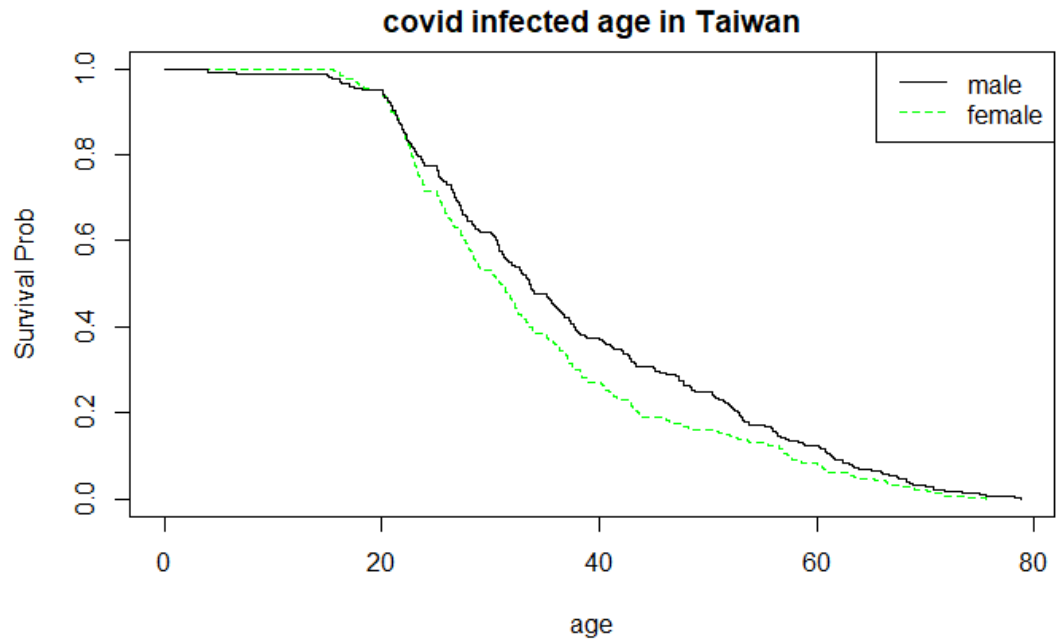

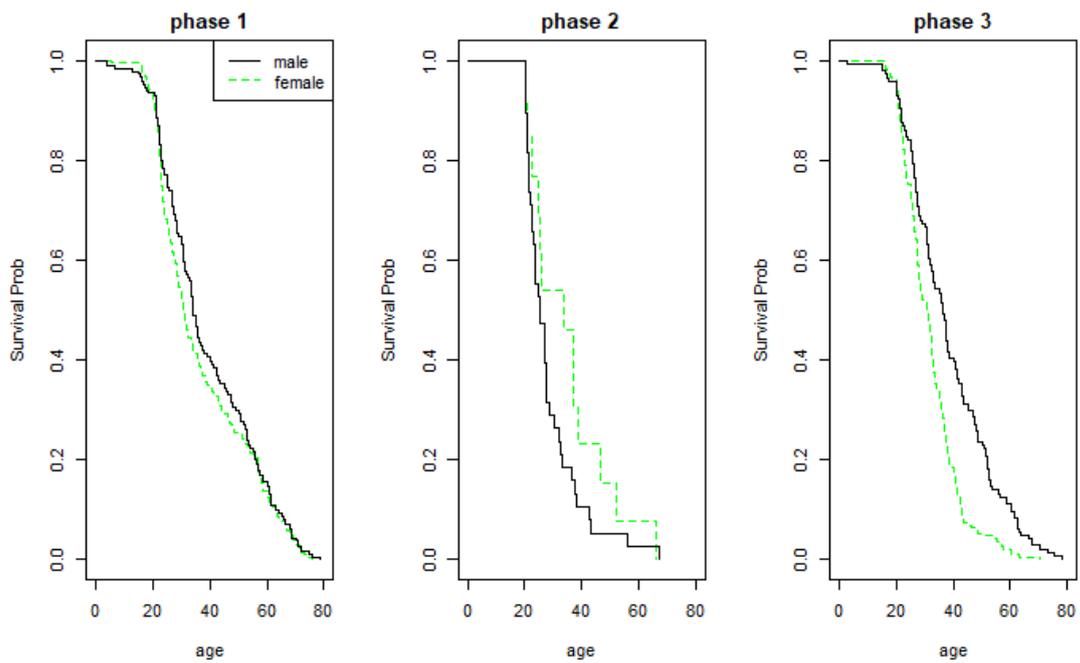

Web Figure 3: The age distributions of male (solid line) and female (dashed line) Covid-19 cases in Taiwan in 2020. Upper: 1/22~12/31; Lower left: Phase 1 (1/22~4/17), Lower middle: Phase 2 (4/18~6/21), Lower right: Phase 3 (6/22~12/31).

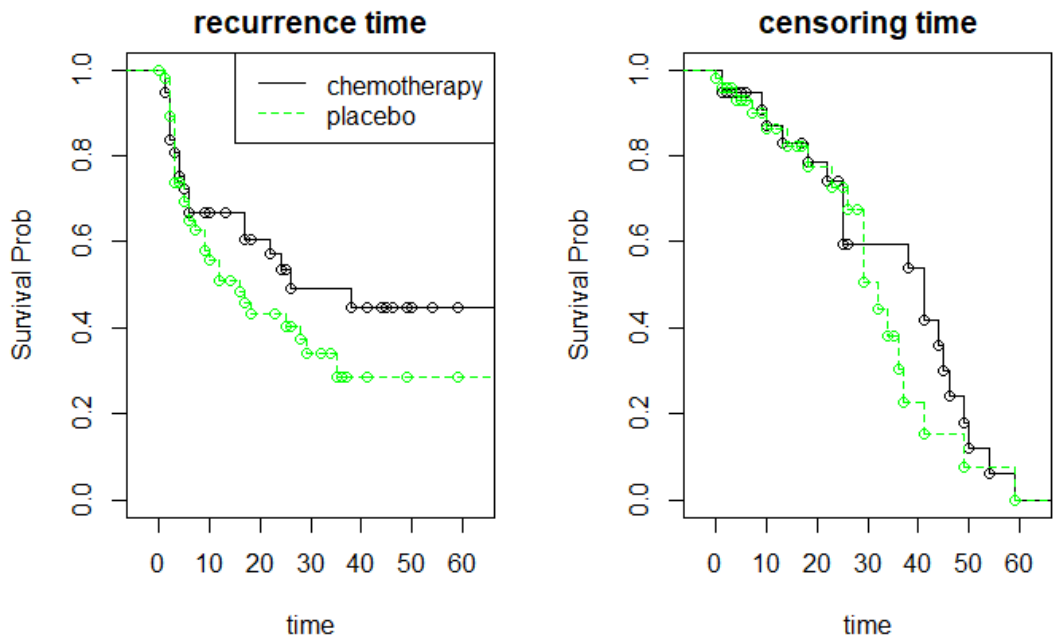

Web Figure 4: Left plot shows the KM curves of the time to first tumor recurrence for patients in the placebo group (Group 0, dotted line) and the chemotherapy group (Group 1, solid line). Right plot shows the KM curves of censoring distributions.

Web Table 1: Summary statistics of $\bar{\tau}_b$ and $\hat{\tau}_b$ based on $n=100$ under the fixed grouping design. For the censored settings, censoring rates for Group 0 and Group 1 are (a) 0.50, 0.09; (b) 0.50, 0.33; (c) 0.5, 0.5; (d) 0.50, 0.67. "Avg. Bias" is the average bias shown in $\times 10^{-3}$ with the standard deviation in parenthesis and "C. Prob." is the coverage probability of 95% confidence intervals with the average length in parenthesis based on 2000 runs.

|  |  | Complete ($\bar{\tau}_b$) |  | Censored ($\hat{\tau}_b$) |  |
| --- | --- | --- | --- | --- | --- |
| Setting | Group Proportion | Avg. Bias (SD) | C. Prob. (Length) | Avg. Bias (SD) | C. Prob. (Length) |
| (a) $\kappa_0 = \kappa_1 = 1$ $\lambda_1 = 10$ $\lambda_0 = 1$ $\tau_b \approx -0.82$ | $p_1 = 0.4$ | 0.237 (0.058) | 0.923 (0.224) | -0.842 (0.060) | 0.934 (0.237) |
|  | $p_1 = 0.5$ | -0.536 (0.061) | 0.915 (0.234) | -0.045 (0.063) | 0.926 (0.246) |
|  | $p_1 = 0.7$ | 0.563 (0.077) | 0.883 (0.283) | -0.339 (0.077) | 0.893 (0.291) |
| (b) $\kappa_0 = \kappa_1 = 1$ $\lambda_1 = 2$ $\lambda_0 = 1$ $\tau_b \approx -0.33$ | $p_1 = 0.4$ | -1.094 (0.108) | 0.942 (0.419) | 1.512 (0.121) | 0.950 (0.486) |
|  | $p_1 = 0.5$ | -3.062 (0.110) | 0.941 (0.420) | 3.911 (0.120) | 0.954 (0.485) |
|  | $p_1 = 0.7$ | -1.889 (0.125) | 0.933 (0.476) | 5.854 (0.142) | 0.937 (0.544) |
| (c) $\kappa_0 = \kappa_1 = 1$ $\lambda_1 = 1$ $\lambda_0 = 1$ $\tau_b = 0$ | $p_1 = 0.4$ | -1.474 (0.118) | 0.950 (0.461) | -0.940 (0.146) | 0.947 (0.572) |
|  | $p_1 = 0.5$ | -3.965 (0.118) | 0.944 (0.452) | 2.119 (0.142) | 0.949 (0.561) |
|  | $p_1 = 0.7$ | -2.451 (0.128) | 0.936 (0.491) | 3.594 (0.159) | 0.925 (0.604) |
| (d) $\kappa_0 = \kappa_1 = 1$ $\lambda_1 = 0.5$ $\lambda_0 = 1$ $\tau_b \approx 0.33$ | $p_1 = 0.4$ | -1.414 (0.112) | 0.944 (0.439) | -1.228 (0.153) | 0.936 (0.591) |
|  | $p_1 = 0.5$ | -3.595 (0.110) | 0.943 (0.422) | -10.746 (0.145) | 0.936 (0.573) |
|  | $p_1 = 0.7$ | -2.134 (0.113) | 0.934 (0.437) | -12.122 (0.157) | 0.935 (0.597) |

Web Table 2: Summary statistics of $\bar{\tau}_b$ and $\hat{\tau}_b$ based on $n=100$ under the random grouping design. For the censored settings, censoring rates for Group 0 and Group 1 are (a) 0.50, 0.09; (b) 0.50, 0.33; (c) 0.5, 0.5; (d) 0.50, 0.67. "Avg. Bias" is the average bias shown in $\times 10^{-3}$ with the standard deviation in parenthesis and "C. Prob." is the coverage probability of 95% confidence intervals with the average length in parenthesis based on 2000 runs.

| Setting | Group Proportion | Complete ($\bar{\tau}_b$) Avg. Bias (SD) | Complete ($\bar{\tau}_b$) C. Prob. (Length) | Censored ($\hat{\tau}_b$) Avg. Bias (SD) | Censored ($\hat{\tau}_b$) C. Prob. (Length) |
|---|---|---|---|---|---|
| (a) $\kappa_0 = \kappa_1 = 1$ $\lambda_1 = 10$ $\lambda_0 = 1$ $\tau_b \approx -0.82$ | $p_1 = 0.4$ | -1.415 (0.060) | 0.909 (0.223) | 2.333 (0.060) | 0.936 (0.239) |
| | $p_1 = 0.5$ | -0.640 (0.064) | 0.895 (0.234) | 2.233 (0.062) | 0.932 (0.248) |
| | $p_1 = 0.7$ | 0.924 (0.076) | 0.878 (0.284) | -0.126 (0.079) | 0.890 (0.293) |
| (b) $\kappa_0 = \kappa_1 = 1$ $\lambda_1 = 2$ $\lambda_0 = 1$ $\tau_b \approx -0.33$ | $p_1 = 0.4$ | -3.394 (0.109) | 0.942 (0.421) | 1.342 (0.123) | 0.940 (0.486) |
| | $p_1 = 0.5$ | -1.580 (0.111) | 0.938 (0.423) | 0.853 (0.125) | 0.944 (0.486) |
| | $p_1 = 0.7$ | 3.380 (0.127) | 0.928 (0.480) | -1.340 (0.140) | 0.934 (0.546) |
| (c) $\kappa_0 = \kappa_1 = 1$ $\lambda_1 = 1$ $\lambda_0 = 1$ $\tau_b = 0$ | $p_1 = 0.4$ | -3.774 (0.119) | 0.940 (0.463) | 0.249 (0.147) | 0.939 (0.574) |
| | $p_1 = 0.5$ | -2.250 (0.117) | 0.945 (0.455) | -0.530 (0.146) | 0.936 (0.564) |
| | $p_1 = 0.7$ | 3.268 (0.131) | 0.934 (0.495) | -4.166 (0.159) | 0.935 (0.606) |
| (d) $\kappa_0 = \kappa_1 = 1$ $\lambda_1 = 0.5$ $\lambda_0 = 1$ $\tau_b \approx 0.33$ | $p_1 = 0.4$ | -2.659 (0.113) | 0.943 (0.441) | -11.759 (0.153) | 0.940 (0.592) |
| | $p_1 = 0.5$ | -2.123 (0.108) | 0.949 (0.424) | -10.688 (0.148) | 0.941 (0.574) |
| | $p_1 = 0.7$ | 2.751 (0.116) | 0.930 (0.439) | -20.334 (0.162) | 0.938 (0.599) |

Web Table 3: Summary statistics of $\bar{\tau}_b$ and $\hat{\tau}_b$ based on $n = 200$ under the fixed grouping design. For the censored settings, censoring rates for Group 0 and Group 1 are (a) 0.50, 0.09; (b) 0.50, 0.33; (c) 0.5, 0.5; (d) 0.50, 0.67. "Avg. Bias" is the average bias shown in $\times 10^{-3}$ with the standard deviation in parenthesis and "C. Prob." is the coverage probability of 95% confidence intervals with the average length in parenthesis based on 2000 runs.

| Setting | Group Proportion | Complete ($\bar{\tau}_b$) Avg. Bias (SD) | C. Prob. (Length) | Censored ($\hat{\tau}_b$) Avg. Bias (SD) | C. Prob. (Length) |
|---|---|---|---|---|---|
| (a) $\kappa_0 = \kappa_1 = 1$ $\lambda_1 = 10$ $\lambda_0 = 1$ $\tau_b \approx -0.82$ | $p_1 = 0.4$ | -2.573 (0.040) | 0.936 (0.158) | 0.680 (0.043) | 0.943 (0.167) |
| | $p_1 = 0.5$ | -3.060 (0.042) | 0.937 (0.166) | 0.412 (0.045) | 0.933 (0.174) |
| | $p_1 = 0.7$ | -3.479 (0.053) | 0.921 (0.202) | -1.136 (0.056) | 0.920 (0.209) |
| (b) $\kappa_0 = \kappa_1 = 1$ $\lambda_1 = 2$ $\lambda_0 = 1$ $\tau_b \approx -0.33$ | $p_1 = 0.4$ | -3.265 (0.075) | 0.947 (0.296) | 1.241 (0.087) | 0.947 (0.340) |
| | $p_1 = 0.5$ | -3.549 (0.075) | 0.946 (0.297) | 0.476 (0.087) | 0.947 (0.340) |
| | $p_1 = 0.7$ | -4.875 (0.087) | 0.941 (0.337) | -0.233 (0.100) | 0.935 (0.384) |
| (c) $\kappa_0 = \kappa_1 = 1$ $\lambda_1 = 1$ $\lambda_0 = 1$ $\tau_b = 0$ | $p_1 = 0.4$ | -2.647 (0.083) | 0.949 (0.327) | 0.488 (0.103) | 0.943 (0.403) |
| | $p_1 = 0.5$ | -2.638 (0.081) | 0.945 (0.320) | -0.435 (0.101) | 0.946 (0.395) |
| | $p_1 = 0.7$ | -4.104 (0.089) | 0.943 (0.348) | -0.609 (0.113) | 0.934 (0.429) |
| (d) $\kappa_0 = \kappa_1 = 1$ $\lambda_1 = 0.5$ $\lambda_0 = 1$ $\tau_b \approx 0.33$ | $p_1 = 0.4$ | -1.883 (0.079) | 0.941 (0.311) | -1.867 (0.108) | 0.942 (0.417) |
| | $p_1 = 0.5$ | -1.881 (0.076) | 0.945 (0.298) | -3.299 (0.103) | 0.948 (0.404) |
| | $p_1 = 0.7$ | -2.805 (0.078) | 0.942 (0.309) | -8.459 (0.111) | 0.940 (0.425) |

Web Table 4: Summary statistics of $\bar{\tau}_b$ and $\hat{\tau}_b$ based on $n=200$ under the random grouping design. For the censored settings, censoring rates for Group 0 and Group 1 are (a) 0.50, 0.09; (b) 0.50, 0.33; (c) 0.5, 0.5; (d) 0.50, 0.67. "Avg. Bias" is the average bias shown in $\times 10^{-3}$ with the standard deviation in parenthesis and "C. Prob." is the coverage probability of 95% confidence intervals with the average length in parenthesis based on 2000 runs.

| Setting | Group Proportion | Complete ($\bar{\tau}_b$) Avg. Bias (SD) | Complete ($\bar{\tau}_b$) C. Prob. (Length) | Censored ($\hat{\tau}_b$) Avg. Bias (SD) | Censored ($\hat{\tau}_b$) C. Prob. (Length) |
|---|---|---|---|---|---|
| (a) $\kappa_0 = \kappa_1 = 1$ $\lambda_1 = 10$ $\lambda_0 = 1$ $\tau_b \approx -0.82$ | $p_1 = 0.4$ | -1.354 (0.041) | 0.933 (0.158) | 0.178 (0.042) | 0.936 (0.167) |
| | $p_1 = 0.5$ | -1.166 (0.043) | 0.926 (0.167) | -0.327 (0.044) | 0.931 (0.174) |
| | $p_1 = 0.7$ | -0.122 (0.055) | 0.908 (0.205) | -0.735 (0.055) | 0.921 (0.210) |
| (b) $\kappa_0 = \kappa_1 = 1$ $\lambda_1 = 2$ $\lambda_0 = 1$ $\tau_b \approx -0.33$ | $p_1 = 0.4$ | -1.937 (0.076) | 0.950 (0.298) | 2.119 (0.087) | 0.943 (0.342) |
| | $p_1 = 0.5$ | -1.435 (0.076) | 0.948 (0.298) | 1.568 (0.087) | 0.949 (0.341) |
| | $p_1 = 0.7$ | -0.825 (0.088) | 0.935 (0.340) | -0.161 (0.098) | 0.947 (0.384) |
| (c) $\kappa_0 = \kappa_1 = 1$ $\lambda_1 = 1$ $\lambda_0 = 1$ $\tau_b = 0$ | $p_1 = 0.4$ | -1.962 (0.084) | 0.952 (0.327) | 1.199 (0.104) | 0.940 (0.404) |
| | $p_1 = 0.5$ | -1.199 (0.083) | 0.944 (0.321) | 0.738 (0.101) | 0.946 (0.396) |
| | $p_1 = 0.7$ | -0.318 (0.090) | 0.946 (0.350) | -0.197 (0.109) | 0.953 (0.429) |
| (d) $\kappa_0 = \kappa_1 = 1$ $\lambda_1 = 0.5$ $\lambda_0 = 1$ $\tau_b \approx 0.33$ | $p_1 = 0.4$ | -1.758 (0.080) | 0.951 (0.312) | -4.645 (0.107) | 0.945 (0.418) |
| | $p_1 = 0.5$ | -1.029 (0.077) | 0.946 (0.299) | -5.080 (0.104) | 0.946 (0.405) |
| | $p_1 = 0.7$ | 0.792 (0.080) | 0.943 (0.310) | -7.361 (0.109) | 0.953 (0.425) |

Web Table 5: Summary statistics of $\bar{\tau}_b$ and $\hat{\tau}_b$ based on $n=400$ under the fixed grouping design. For the censored settings, censoring rates for Group 0 and Group 1 are (a) 0.50, 0.09; (b) 0.50, 0.33; (c) 0.5, 0.5; (d) 0.50, 0.67. "Avg. Bias" is the average bias shown in $\times 10^{-3}$ with the standard deviation in parenthesis and "C. Prob." is the coverage probability of 95% confidence intervals with the average length in parenthesis based on 2000 runs.

|  |  | Complete ($\bar{\tau}_b$) |  | Censored ($\hat{\tau}_b$) |  |
| --- | --- | --- | --- | --- | --- |
| Setting | Group Proportion | Avg. Bias (SD) | C. Prob. (Length) | Avg. Bias (SD) | C. Prob. (Length) |
| (a) $\kappa_0 = \kappa_1 = 1$ $\lambda_1 = 10$ $\lambda_0 = 1$ $\tau_b \approx -0.82$ | $p_1 = 0.4$ | 0.035 (0.029) | 0.944 (0.113) | -0.750 (0.030) | 0.941 (0.117) |
|  | $p_1 = 0.5$ | 0.253 (0.030) | 0.943 (0.119) | -0.793 (0.031) | 0.947 (0.123) |
|  | $p_1 = 0.7$ | -0.201 (0.037) | 0.934 (0.145) | -1.178 (0.038) | 0.933 (0.150) |
| (b) $\kappa_0 = \kappa_1 = 1$ $\lambda_1 = 2$ $\lambda_0 = 1$ $\tau_b \approx -0.33$ | $p_1 = 0.4$ | 1.668 (0.054) | 0.946 (0.210) | -0.588 (0.061) | 0.952 (0.240) |
|  | $p_1 = 0.5$ | 1.900 (0.055) | 0.941 (0.211) | -1.035 (0.060) | 0.952 (0.240) |
|  | $p_1 = 0.7$ | 1.249 (0.061) | 0.945 (0.240) | -2.280 (0.068) | 0.947 (0.271) |
| (c) $\kappa_0 = \kappa_1 = 1$ $\lambda_1 = 1$ $\lambda_0 = 1$ $\tau_b = 0$ | $p_1 = 0.4$ | 2.323 (0.060) | 0.946 (0.231) | -1.341 (0.072) | 0.954 (0.285) |
|  | $p_1 = 0.5$ | 2.787 (0.059) | 0.941 (0.226) | -1.926 (0.070) | 0.949 (0.279) |
|  | $p_1 = 0.7$ | 1.953 (0.064) | 0.949 (0.247) | -3.606 (0.077) | 0.946 (0.304) |
| (d) $\kappa_0 = \kappa_1 = 1$ $\lambda_1 = 0.5$ $\lambda_0 = 1$ $\tau_b \approx 0.33$ | $p_1 = 0.4$ | 2.728 (0.058) | 0.941 (0.219) | -3.597 (0.074) | 0.959 (0.296) |
|  | $p_1 = 0.5$ | 3.280 (0.055) | 0.939 (0.210) | -4.520 (0.073) | 0.954 (0.286) |
|  | $p_1 = 0.7$ | 2.151 (0.057) | 0.944 (0.219) | -8.563 (0.077) | 0.950 (0.304) |

Web Table 6: Summary statistics of $\bar{\tau}_b$ and $\hat{\tau}_b$ based on $n=400$ with unequal censoring distributions. For the censored settings, censoring rates for Group 0 and Group 1 are (a) 0.50, 0.09; (b) 0.50, 0.33; (c) 0.5, 0.5; (d) 0.50, 0.67. "Avg. Bias" is the average bias shown in $\times 10^{-3}$ with the standard deviation in parenthesis and "C. Prob." is the coverage probability of 95% confidence intervals with the average length in parenthesis based on 2000 runs.

| Setting | Group Proportion | Fixed grouping Avg. Bias (SD) | Fixed grouping C. Prob. (Length) | Random grouping Avg. Bias (SD) | Random grouping C. Prob. (Length) |
|---|---|---|---|---|---|
| (a) $\kappa_0 = \kappa_1 = 1$ $\lambda_1 = 10$ $\lambda_0 = 1$ $\tau_b \approx -0.82$ | $p_1 = 0.4$ | -1.057 (0.029) | 0.943 (0.116) | -0.245 (0.029) | 0.950 (0.116) |
| | $p_1 = 0.5$ | -1.030 (0.031) | 0.945 (0.122) | -0.679 (0.031) | 0.944 (0.122) |
| | $p_1 = 0.7$ | -1.337 (0.378) | 0.934 (0.149) | -1.495 (0.039) | 0.931 (0.149) |
| (b) $\kappa_0 = \kappa_1 = 1$ $\lambda_1 = 2$ $\lambda_0 = 1$ $\tau_b \approx -0.33$ | $p_1 = 0.4$ | -1.602 (0.058) | 0.948 (0.230) | -1.792 (0.059) | 0.945 (0.230) |
| | $p_1 = 0.5$ | -2.040 (0.059) | 0.945 (0.232) | -2.475 (0.060) | 0.941 (0.232) |
| | $p_1 = 0.7$ | -2.976 (0.067) | 0.954 (0.266) | -1.875 (0.070) | 0.938 (0.266) |
| (c) $\kappa_0 = \kappa_1 = 1$ $\lambda_1 = 1$ $\lambda_0 = 1$ $\tau_b = 0$ | $p_1 = 0.4$ | -1.739 (0.067) | 0.949 (0.266) | -1.644 (0.069) | 0.944 (0.266) |
| | $p_1 = 0.5$ | -2.148 (0.067) | 0.949 (0.264) | -2.247 (0.069) | 0.941 (0.264) |
| | $p_1 = 0.7$ | -3.492 (0.075) | 0.942 (0.294) | -1.772 (0.078) | 0.934 (0.293) |
| (d) $\kappa_0 = \kappa_1 = 1$ $\lambda_1 = 0.5$ $\lambda_0 = 1$ $\tau_b \approx 0.33$ | $p_1 = 0.4$ | -4.881 (0.067) | 0.953 (0.268) | -4.274 (0.069) | 0.944 (0.268) |
| | $p_1 = 0.5$ | -5.329 (0.066) | 0.949 (0.263) | -4.820 (0.068) | 0.941 (0.263) |
| | $p_1 = 0.7$ | -8.155 (0.073) | 0.950 (0.287) | -6.103 (0.077) | 0.936 (0.285) |

Web Table 7: Power comparison for the LR, Gehan and proposed $\hat{\tau}_b$ tests for testing $H_0: S_0 = S_1$ under the PH assumption and three configurations of censoring distributions.

(a) $(\lambda_0^C, \lambda_1^C) = (1,1)$

| $\tau_b$ | 0 | 0.032 | 0.067 | 0.103 | 0.143 | 0.185 | 0.231 | 0.280 | 0.333 |
|---|---|---|---|---|---|---|---|---|---|
| $\lambda_1/\lambda_0$ | 1 | 0.938 | 0.875 | 0.813 | 0.750 | 0.688 | 0.625 | 0.563 | 0.500 |
| Power: LR | 0.057 | 0.079 | 0.153 | 0.299 | 0.484 | 0.703 | 0.865 | 0.957 | 0.994 |
| Power: Gehan | 0.545 | 0.077 | 0.137 | 0.246 | 0.394 | 0.588 | 0.760 | 0.894 | 0.969 |
| Power: Proposed | 0.060 | 0.082 | 0.154 | 0.300 | 0.481 | 0.698 | 0.864 | 0.951 | 0.988 |
| $\Pr(\delta=0\mid X=0)$ | 0.5 | 0.5 | 0.5 | 0.5 | 0.5 | 0.5 | 0.5 | 0.5 | 0.5 |
| $\Pr(\delta=0\mid X=1)$ | 0.501 | 0.517 | 0.534 | 0.552 | 0.572 | 0.592 | 0.615 | 0.640 | 0.666 |

(b) $(\lambda_0^C, \lambda_1^C) = (0.2, 0.2)$

| $\tau_b$ | 0 | 0.032 | 0.067 | 0.103 | 0.143 | 0.185 | 0.231 | 0.280 | 0.333 |
|---|---|---|---|---|---|---|---|---|---|
| Power: LR | 0.061 | 0.101 | 0.221 | 0.467 | 0.733 | 0.906 | 0.983 | 1.000 | 1.000 |
| Power: Gehan | 0.055 | 0.093 | 0.191 | 0.378 | 0.609 | 0.821 | 0.949 | 0.992 | 1.000 |
| Power: Proposed | 0.057 | 0.097 | 0.208 | 0.419 | 0.666 | 0.870 | 0.973 | 0.997 | 1.000 |
| $\Pr(\delta=0\mid X=0)$ | 0.166 | 0.166 | 0.166 | 0.166 | 0.166 | 0.166 | 0.166 | 0.166 | 0.166 |
| $\Pr(\delta=0\mid X=1)$ | 0.166 | 0.175 | 0.185 | 0.196 | 0.209 | 0.224 | 0.242 | 0.262 | 0.286 |

(c) $(\lambda_0^C, \lambda_1^C) = (0.05, 0.05)$

| $\tau_b$ | 0 | 0.032 | 0.067 | 0.103 | 0.143 | 0.185 | 0.231 | 0.280 | 0.333 |
|---|---|---|---|---|---|---|---|---|---|
| Power: LR | 0.054 | 0.102 | 0.248 | 0.516 | 0.792 | 0.947 | 0.992 | 0.999 | 1.000 |
| Power: Gehan | 0.058 | 0.091 | 0.214 | 0.422 | 0.664 | 0.873 | 0.973 | 0.998 | 1.000 |
| Power: Proposed | 0.060 | 0.095 | 0.222 | 0.437 | 0.684 | 0.886 | 0.978 | 0.998 | 1.000 |
| $\Pr(\delta=0\mid X=0)$ | 0.047 | 0.047 | 0.047 | 0.047 | 0.047 | 0.047 | 0.047 | 0.047 | 0.047 |
| $\Pr(\delta=0\mid X=1)$ | 0.047 | 0.050 | 0.054 | 0.057 | 0.062 | 0.067 | 0.073 | 0.081 | 0.090 |

Web Table 8: Summary statistics of the standardized LR, Gehan and proposed $\hat{\tau}_b$ statistics under three censoring settings when the PH assumption does not hold.

| | | Min. | 1st Qu. | Median | Mean | 3rd Qu. | Max. |
|---|---|---|---|---|---|---|---|
| (a) | LR | -4.473 | -1.916 | -1.210 | -1.222 | -0.538 | 1.901 |
| $\Pr(\delta=0\mid X=0)=0.604$ | Gehan | -8.058 | -5.949 | -5.346 | -5.337 | -4.763 | -2.356 |
| $\Pr(\delta=0\mid X=1)=0.560$ | Proposed | -3.434 | -0.597 | 0.032 | 0.052 | 0.712 | 3.527 |
| (b) | LR | -3.661 | -0.562 | 0.145 | 0.117 | 0.821 | 3.595 |
| $\Pr(\delta=0\mid X=0)=0.490$ | Gehan | -7.776 | -5.161 | -4.506 | -4.503 | -3.880 | -1.020 |
| $\Pr(\delta=0\mid X=1)=0.506$ | Proposed | -2.965 | -0.534 | 0.107 | 0.092 | 0.778 | 3.430 |
| (c) | LR | -2.895 | 0.562 | 1.246 | 1.227 | 1.917 | 4.405 |
| $\Pr(\delta=0\mid X=0)=0.390$ | Gehan | -7.390 | -4.336 | -3.651 | -3.673 | -3.016 | -0.048 |
| $\Pr(\delta=0\mid X=1)=0.453$ | Proposed | -3.256 | -0.570 | 0.136 | 0.108 | 0.779 | 3.284 |

Web Table 9A: Summary statistics for Taiwan's COVID-19 data in 2020

|  | overall | Phase 1 | Phase 2 | Phase 3 |
|---|---|---|---|---|
| $n$ | 780 | 395 | 51 | 334 |
| $N_0$ | 414 | 211 | 13 | 190 |
| $N_1$ | 366 | 184 | 38 | 144 |
| Median age (female) | 30.91 | 30.93 | 33.53 | 30.85 |
| Median age (male) | 33.62 | 33.95 | 25.22 | 36.59 |
| $\bar{\tau}_b$ | 0.114 | 0.087 | -0.287 | 0.257 |
| C.I. (random) | (0.033, 0.195) | (-0.027, 0.200) | (-0.651, 0.077) | (0.133, 0.380) |
| $H_0^{tau}: \tau_b = 0$ | 0.0068 | 0.1404 | 0.1658 | 0.0002 |

Web Table 9B: Number of female and male confirmed cases in different age categories based on Taiwan's COVID-19 data in 2020 (p-value of Chi-squared test = 0.0456)

| Age Group | 0-19 | 20-24 | 25-29 | 30-34 | 35-39 | 40-44 | 45-49 | 50-54 | 55-59 | 60-64 | 65-69 | 70+ |
|---|---|---|---|---|---|---|---|---|---|---|---|---|
| Female | 20 | 97 | 76 | 62 | 47 | 33 | 13 | 12 | 20 | 15 | 10 | 9 |
| Male | 18 | 64 | 57 | 52 | 38 | 25 | 21 | 28 | 17 | 21 | 13 | 12 |

Web Table 10A: The values of $\hat{\tau}_b^*$ statistic with the tail after $t^*$ imputed by four parametric models for the bladder cancer example.

| Imputed parametric model | $t^* = 30$ | $t^* = 40$ | $t^* = 50$ | $t^* = 59$ |
|---|---|---|---|---|
| Exponential | 0.173 | 0.168 | 0.157 | 0.152 |
| Weibull | 0.197 | 0.189 | 0.176 | 0.167 |
| Log-normal | 0.184 | 0.182 | 0.173 | 0.167 |
| Logistic | 0.190 | 0.165 | 0.151 | 0.147 |

Web Table 10B: Bootstrapped confidence intervals based on $\hat{\tau}_b^*$ and $\hat{\tau}_b^R$ statistics with the tail after $t^*$ imputed by four parametric models for the bladder cancer example.

| $\hat{\tau}_b^*$ | C.I. (random) | C.I. (fixed) |
|---|---|---|
| Exponential | (-0.119, 0.420) | (-0.116, 0.408) |
| Weibull | (-0.114, 0.437) | (-0.110, 0.424) |
| Log-normal | (-0.116, 0.440) | (-0.112, 0.431) |
| Logistic | (-0.117, 0.410) | (-0.113, 0.399) |
| $\hat{\tau}_b^R$ | (-0.136, 0.627) | (-0.126, 0.595) |

Web Table 10C: P values for testing $H_0 : S_0 = S_1$ for the bladder cancer example.

|  | LR | Gehan | $\hat{\tau}_b$ |
|---|---|---|---|
| p-value | 0.20 | 0.39 | 0.29 |

**ADDITIONAL REFERENCES**